\documentclass[aps,prd,twocolumn,groupedaddress,nofootinbib]{revtex4}

\pdfoutput=1
\usepackage {amsmath}
\usepackage {amssymb}
\usepackage {amsfonts}
\usepackage {amsthm} 
\usepackage {mathrsfs}
\usepackage {natbib}
\usepackage {latexsym}
\usepackage {graphicx}
\usepackage {dsfont}
\usepackage {txfonts}
\usepackage {rotating}
\usepackage {wasysym}
\usepackage {multirow}
\usepackage {hhline}
\usepackage {graphicx}
\usepackage {hyperref}
\usepackage {color}
\usepackage {bm}
\usepackage{appendix}
\usepackage{acronym}

\newcommand{\dcc}{LIGO-P1000020-v2}

\def\commitID{commitID: aa83996fd7c01dd8636c284a174d657577b169c9}
\def\commitDATE{Wed Aug 3 15:32:27 2011 +0100}

\begin{document}

\title{A semi-coherent search strategy for known
  continuous wave sources in binary systems}
\author{C.~Messenger}
\email{chris.messenger@astro.cf.ac.uk}
\affiliation{School of Physics and Astronomy, Cardiff University,
  Queens Buildings, The Parade, Cardiff, CF24 3AA}
\date{\today}

\date{\commitDATE\\\mbox{\small \commitID}\\\mbox{\dcc}}

\begin{abstract}
  We present a method for detection of weak continuous signals from
  sources in binary systems via the incoherent combination of many
  ``short'' coherently-analyzed segments.  The main focus of the work
  is on the construction of a metric on the parameter space for such
  signals for use in matched-filter based searches.  The metric is
  defined using a maximum likelihood detection statistic applied to a
  binary orbit phase model including eccentricity.  We find that this
  metric can be accurately approximated by its diagonal form in the
  regime where the segment length is $\ll$ the orbital period.  Hence
  correlations between parameters are effectively removed by the
  combination of many independent observation.  We find that the
  ability to distinguish signal parameters is independent of the total
  semi-coherent observation span (for the semi-coherent span $\gg$ the
  segment length) for all but the orbital angular frequency.
  Increased template density for this parameter scales linearly with
  the observation span.  We also present two example search schemes.
  The first uses a re-parameterized phase model upon which we compute
  the metric on individual short coherently analyzed segments.  The
  second assumes long $\gg$ the orbital period segment lengths from
  which we again compute the coherent metric and find it to be
  approximately diagonal.  In this latter case we also show that the
  semi-coherent metric is equal to the coherent metric.
\end{abstract}

\maketitle

\acrodef{GW}[GW]{gravitational-wave}
\acrodef{NS}[NS]{neutron star}
\acrodef{EM}[EM]{electromagnetic}
\acrodef{SNR}[SNR]{signal-to-noise-ratio}
\acrodef{LIGO}[LIGO]{Laser Interferometer Gravitational-wave
  Observatory}
\acrodef{LMXB}[LMXB]{low-mass X-ray binary}
\acrodefplural{LMXB}[LMXBs]{low-mass X-ray binaries}

\section{Introduction\label{sec:introduction}}

The search for continuously radiating sources in binary systems has
proven to be a consistently intensive endeavor in a number of
branches of astrophysics.  The prime focus being the search for
pulsars in the fields of Radio and X-ray astronomy, and more recently
for non-axisymmetric rapidly rotating neutron-stars in \ac{GW}
astronomy.  In the Radio and X-ray fields it has long been known how
to perform searches for such systems over relatively short observation
times ($\ll$ orbital period), known as ``acceleration''
searches~\cite{1990Natur.346...42A,1992PhDT........20A,2004MNRAS.355..147F}.
In this type of search the frequency of a Doppler modulated signal
from a source in a binary system is, over the short observation,
approximated by its Taylor expansion and the search is performed over
a parameter space defined by the frequency and its derivatives.  More
recent searches, sensitive in the complementary extreme to systems
with orbital periods $\ll$ the observation time, have also been
successful in the detection of pulsars in the Radio
band~\cite{2003ApJ...589..911R,2007ApJ...670..363H,2005ASPC..328..199R}.
These searches take advantage of the distinct frequency domain
signature generated by a frequency modulated signal such as a Doppler
modulated continuously emitting pulsar in a binary system.  This type
of search is also planned for application to the \ac{GW}
case~\cite{2007CQGra..24..469M}.

In this work we are motivated by the problems inherent to detection of
\ac{GW} radiation from known binary systems, of which primary examples
are the \acp{LMXB}.  In this case the signal is expected to be
extremely weak and the parameter space known to be very large, the
proverbial ``needle in a haystack''.  This is also a long standing
issue for X-ray astronomy for a subset of these
objects~\cite{1991ApJ...379..295W,1994ApJ...435..362V}.  Here we
consider known objects as systems where the sky-position is assumed to
be known precisely and that the intrinsic source frequency, as well as
the orbital parameters of the system, are unknown.  In the search for
\acp{GW} from such objects, to date, two strategies have been
employed, one being a fully coherent
analysis~\cite{2007PhRvD..76h2001A} spanning only a short observation
time, and the other employing a cross-correlation
technique~\cite{2007PhRvD..76h2003A} using data from multiple
detectors.  In the former a parameter space covering was used based on
the coherent parameter space metric on a non-eccentric phase model
with known orbital period~\cite{2001PhRvD..63l2001D}.  The coherent
metric, as we will show in the Section~\ref{sec:coherentmetric}, is
simply a measure of ``distance'' defined on the parameter space and
informs us on how to place templates (or filters) optimally within the
parameter space for a coherent search.  Here we aim to expand on this
approach and compute the semi-coherent metric, a similar measure of
distance but defined on a semi-coherent detection statistic.
 
The first work on the coherent parameter space metric for sources in
binary systems~\cite{2001PhRvD..63l2001D} has recently been built upon
by~\cite{2008MNRAS.389..839W} where it has been shown for the \ac{GW}
case that searches for the \acp{LMXB} are computationally bound
meaning that the number of templates one is required to process to
optimally cover the parameter space is too large to computed on human
time-scales.  The apparently optimal fully-coherent approach is
therefore unfeasible.  This is a common theme within continuous
\ac{GW} data analysis for large parameter space searches and has been
addressed via various applications of what we will call
``semi-coherent'' searches.  This is where an observation is divided
into a number of individual coherent observations, or ``segments'',
which are then incoherently combined to form a more powerful detection
statistic.  The term ``semi-coherent'' is in reference to the fact
that the unknown phase of the signal has been either maximized or
marginalized away within each segment of the analysis and hence there
is no required phase coherence between the signal and our phase model
for the duration of the entire observation span.  Since the number of
templates required for fully-coherent searches typically scale with
the coherent observation time to a power $l>1$, by dividing up the
observation into $M$ segments the number of templates required to
cover all segments (the number for each segment multiplied by the
number of segments) will be reduced by a factor of $M^{l-1}$.  The
missing piece of the puzzle for sources in binary systems is the
method by which to combine the detection statistics computed on the
templates within each coherently analyzed segment.  This is also true
in the case of the search for X-ray pulsations from the \acp{LMXB}
where, in addition to being computationally bound just as in the
\ac{GW} case, single observations of individual sources from X-ray
timing satellites such as RXTE (Rossi X-Ray Timing Explorer) are
typically restricted to time-scales of order $\sim 1$ hour.  This
limitation, due to scheduling constraints and obscuration of the
source as the detector orbits the earth, represents an automatic and
unavoidable division of the complete dataset into segments.

The work presented here is based on that
of~\cite{bala96:_gravit_binaries_metric,owen96:_search_templates,1999PhRvD..60b2002O}
who were the first to describe the concept of the metric, based on a
distance measure equivalent to the expected loss in the detection
statistic between signal and template (the mismatch) in a
matched-filter based search.  From such a metric one would then
proceed to place templates within the parameter space based on the
criteria that the mismatch at any given point was always less than a
desired threshold.  In~\cite{2000PhRvD..61h2001B} is was shown that
the metric on the general semi-coherent detection statistic was simply
the average of the coherent metrics from the constituent segments.
Recent work by~\cite{2008PhRvD..78j2005P,2009PhRvL.103r1102P} has
described the construction of such a metric on the parameter space
defined by the frequency and it's derivatives and the sky coordinates.
This result is specific to all-sky, wide band, searches for \acp{GW}
from unknown isolated sources.  In this work we compute the
semi-coherent metric for the complementary class of sources, the known
sources with known sky position in binary systems with unknown
frequency and orbital parameters.

In the remainder of this section we review the binary system signal
model and the power-like detection statistic as well as the coherent
and semi-coherent metric definitions.  In Section~\ref{sec:results} we
describe the calculation of the coherent and semi-coherent metric for
the specific case of sources in binary systems.  In particular we
analyze two complementary scenarios, in
Section~\ref{sec:shortcoherent} we deal with cases where the segment
length is $\ll$ the orbital period of the source and in
Section~\ref{sec:longcoherent} the case where the segment length
$\gg$ the orbital period.  In each case we outline a basic method for
the practical implementation of a search, and in
Section~\ref{sec:discussion} we summarize our findings.
%
\subsection{The signal model\label{sec:signalmodel}}

We model a continuously emitting source of radiation (either \ac{GW}
or \ac{EM}) located
within a binary system with a non-emitting companion.  In this sense
we define the noise-free continuous signal received at an inertial
reference frame to be
\begin{equation}\label{eq:signalmodel}
  s(t) = A\sin\left[\Phi(t) +  \Phi_{0}\right]
\end{equation}
where $A$ as the constant signal amplitude and $\Phi_{0}$ is a
constant phase offset.  Note that in the following we are assuming
that the intrinsic source frequency is constant (i.e. there is no
\emph{intrinsic} frequency-evolution of the source) and that any
detector motion relative to the chosen inertial reference frame has
been accounted for.  The latter assumption in practice requires that
the sky-position of the source is known to high accuracy and that the
time-series has been barycentered (usually to the solar-system
barycenter).  We also ignore the slowly varying amplitude response in
the case of a \ac{GW} signal.  The remaining phase contribution is
assumed to be entirely due to the constant intrinsic frequency and the
Roemer delay across the source orbit.

This time-dependent phase for a source in a bound eccentric
(non-relativistic) orbit is~\cite{1976ApJ...205..580B} 
\begin{eqnarray}
   \Phi(t) &=&
   2\pi\nu\bigg\{t-t_{\mathrm{ref}} \label{eq:exactphasemodel}
   \\
   &+& a\left[\sin\omega(\cos{E(t)}-e) +
     \cos\omega\sin{E(t)}\sqrt{1-e^2}\right]\bigg\} \nonumber
\end{eqnarray}
where $\nu$ is the intrinsic signal frequency\footnote{We have assumed
  that the velocity of the binary barycenter is constant relative to
  the inertial reference frame and hence absorbed any Doppler shifts
  into $\nu$.}, $t_{\mathrm{ref}}$ is
some reference time (at which the signal phase is equal to $\Phi_{0}$
in the source frame), $a$ is the orbital semi-major
axis projected along the line of sight and divided by the speed of
light, $\omega$ is the argument of periapse, $e$ is the orbital
eccentricity and $E$ is the eccentric anomaly.  The eccentric anomaly
is defined by the transcendental relation
\begin{equation}\label{eq:eccentricanomaly}
  \frac{2\pi}{P}\left(t-t_{\mathrm{p}}\right)=E - e\sin{E}
\end{equation}
where $P$ is the orbital period (equal to $2\pi/\Omega$ where $\Omega$
is the orbital angular frequency) and $t_{\mathrm{p}}$ is the time of
periapsis (the point of closest approach to the prime focus of
the orbit). 

With regards to the search for continuous \ac{GW} emission from sources in
binary systems, the primary targets (as mentioned in
Section~\ref{sec:introduction}) are the \acp{LMXB}.  These sources are
observed to have highly circularized orbits (see~\cite{2008MNRAS.389..839W} for
detailed descriptions of current orbital parameter estimates) and as
such we have chosen to limit our investigations to low-eccentricity orbits.   
In the low-eccentricity limit ($e\ll 1$) we can adopt the
approximation~\cite{1988ormo.book.....R,2001MNRAS.326..274L}
\begin{eqnarray}
   \Phi(t) &\approx&
   2\pi\nu\Bigg\{t-t_{\mathrm{ref}} \label{eq:loweccentricityphasemodel}\\
   &+&
    a\left[\sin\psi(t)+\frac{\kappa}{2}\sin{2\psi(t)}-\frac{\eta}{2}\cos{2\psi(t)}-\frac{3\eta}{2}\right]\Bigg\},
   \nonumber 
\end{eqnarray}
where we have Taylor-expanded the orbital contribution to the phase in
powers of the eccentricity $e$ up to leading order.  In addition we
have adopted the following parameters, 
\begin{subequations}
\begin{eqnarray}
  \kappa &=& e\cos\omega, \label{eq:kappa} \\
  \eta &=& e\sin\omega, \label{eq:eta}
\end{eqnarray}
\end{subequations}
as replacements for the more physical $e$ and $\omega$
parameters.  This change is motivated by the fact that for
low-eccentric systems strong degeneracies between the argument of
periapse $\omega$ and the time of periapse passage $t_{\mathrm{p}}$
would complicate our analysis.  In addition we also define
\begin{equation}\label{eq:psi}
  \psi(t) = \Omega\left(t-t_\mathrm{asc}\right),
\end{equation}
as the time dependent orbital phase as measured relative to the time
of passage through the ascending node of the orbit $t_{\mathrm{asc}}= t_{\mathrm{p}}-\omega/\Omega$. 

The circular orbit case is a specific instance of the more general
elliptic orbit (Eq.~\ref{eq:exactphasemodel}).  For $e=0$ we can write
the phase model in this case as
\begin{equation}\label{eq:circularphasemodel}
   \Phi(t) =
   2\pi\nu\left[t-t_{\mathrm{ref}}- a\sin\psi(t)\right]
\end{equation}
where the extraneous parameter $\omega$ has been set to $\omega=\pi$
such that, just as for the Taylor-expanded low eccentricity case, the
orbital reference time is the time of passage through the ascending
node of the orbit rather than the, now meaningless, time of periapsis
$t_{\mathrm{p}}$.

For the majority of our primary sources, the \acp{LMXB}, the expected
eccentricities in these systems are relatively low (typically
$>10^{-3}$).  However, despite the apparent current requirement for
circular orbits only, we continue the analysis using the phase defined
for an low-eccentricity orbit (Eq.~\ref{eq:loweccentricityphasemodel})
since the results that follow from this choice can easily be
interpreted for either the eccentric or circular orbit cases.
%
\subsection{The detection statistic\label{sec:detectionstatistic}}

In searching for deterministic signals in noisy data one wishes, in
general, to distinguish between the noise hypothesis and the
signal-plus-noise hypothesis in order to determine the presence of a
signal.  In this case the likelihood-ratio test is optimal in the
Neyman-Pearson sense~\cite{NeymanPearson1933}.
If we assume that our measured time-series data-set consists of our signal and
additive Gaussian noise sampled uniformly at discrete times $t_{j}$
such that
\begin{equation}\label{eq:dataset}
  x(t_{j}) = s(t_{j},\bm{\theta}') + n(t_{j}),
\end{equation}
where $\bm{\theta}'$ is a vector containing the signal parameters.
We can write the likelihood function as
\begin{equation}\label{eq:likelihood}
  L(\bm{\theta}') \propto
  \exp\left\{-\frac{1}{2}\sum_{j=1}^{N}\sum_{k=1}^{N}
    \Big(x_{j}-s_{j}(\bm{\theta}')\Big)  C^{-1}_{jk}
    \Big(x_{k}-s_{k}(\bm{\theta}')\Big)  \right\},
\end{equation}
where $C$ is the noise covariance matrix, $N$ is the number of samples
in our time-series and for simplicity of notation we have replaced $x(t_{j})$ with
$x_{j}$ and $s(t_{j})$ with $s_{j}$.

One can show that twice\footnote{We multiply the
  log-likelihood ratio by 2 so that it is exactly consistent with a
  $\chi^{2}$ distribution with 2 degrees of freedom.} the log-likelihood ratio (the
log of the ratio between the likelihood function defined above and the likelihood
assuming no signal), when analytically maximized over the ``nuisance'' parameters
$A$ and $\Phi_{0}$, becomes
\begin{equation}\label{eq:loglikelihoodratio}
  \Lambda(\bm{\theta}) =
  \frac{4}{S_{\!\text{n}}\Delta{T}}\left|\sum_{j=1}^{N}\Delta t\, x_{j}e^{-i\Phi_{j}(\bm{\theta})}\right|^2
\end{equation}
in the case of ``white'' noise where the covariance matrix
is diagonal\footnote{This is a valid assumption for practical
  purposes since analyses are typically divided into narrow frequency
  regions within which the noise spectral density can be assumed ``white''.}.
Note that the un-primed $\bm{\theta}$ represents the reduced parameter set
after having removed the dependence upon the nuisance parameters and
that we have used $\Delta{t}$ as our sampling time, $\Delta{T}=N\Delta{t}$ as our
time span and $S_{\!\text{n}}$ as the single-sided noise spectral
density (assumed constant over the frequency band of interest). 
This detection statistic has the form of a Fourier power with the
exception that the complex phase, with which we multiply each datum,
contains an orbital phase component in addition to the intrinsic
frequency component.  

The statistical behavior of the log-likelihood ratio detection
statistic in additive Gaussian noise follows that of a $\chi^{2}$
distribution.  In the simplified constant amplitude case described
here the random variable $\Lambda(\bm{\theta})$ is described by a
non-central $\chi^{2}_{2}$ distribution (the subscript indicates the
number of degrees of freedom).  Assuming a set of signal parameters
$\bm{\theta}$ the expectation value of $\Lambda(\bm{\theta})$
evaluated at an offset parameter space location $\bm{\theta}
+\bm{\Delta\theta}$ is then given by
\begin{equation}\label{eq:expectation}
  E\left[\Lambda(\bm{\theta},\bm{\Delta\theta})\right] = 2 + \rho^{2}(\bm{\theta},\bm{0})\left|\sum_{j=1}^{N}e^{i\Delta\Phi_{j}(\bm{\theta},\bm{\Delta\theta})}\right|^{2},
\end{equation}
where $\rho^{2}(\bm{\theta},\bm{0})$ is the optimal signal-to-noise ratio (SNR) given
by 
\begin{equation}
\rho^{2}(\bm{\theta},\bm{0}) = \frac{2}{S_{\!\text{n}}}\sum_{j=1}^{N}\Delta{t}\,s^{2}_{j}(\bm{\theta}).
\end{equation}
Note that we have used
$\Delta\Phi_{j}(\bm{\theta},\bm{\Delta\theta})=\Phi_{j}(\bm{\theta}+\bm{\Delta\theta})-\Phi_{j}(\bm{\theta})$
to represent the phase offset caused by the offset in parameter space
location.  Also note that the second term in Eq.~\ref{eq:expectation}
is equal to the non-centrality parameter governing the $\chi^{2}_{2}$
distribution which becomes the optimal \ac{SNR} squared when the
template and signal parameters are exactly matched.  In our problem we
are faced with a continuum of signal hypotheses defined by the
parameter space spanned by $\bm{\theta}$, which we must somehow sample
from.  The common frequentist strategy in parameter space searches is
to maximize the detection statistic over all unknown parameter values.
For two of these parameters, $A$ and $\Phi_{0}$, we have performed
this maximization analytically.  In the following section we describe
the standard strategy for performing the remaining
maximizations\footnote{We note that strictly speaking, the optimality
  of the standard likelihood-ratio applies only for the
  point-hypothesis case.  It has been shown~\cite{2008arXiv0804.1161S}
  that the optimal statistic in the more general case (including the
  point hypothesis case) where one is faced with a continuum of signal
  hypotheses (defined one some parameter space) is the Bayesian
  statistic, the ``Bayes Factor''.}.

For a deterministically amplitude modulated signal, such as in the
\ac{GW} case, the standard technique is to analytically maximize not
only over the phase and amplitude of the signal but also over the
inclination angle of the source and the polarization angle of the
\ac{GW} wave~\cite{1998PhRvD..58f3001J}.  In such a case the detection
statistic becomes a $\chi^{2}$ statistic with 4 degrees of freedom.

\subsection{The coherent metric\label{sec:coherentmetric}}

The parameter space in our case is defined by the ranges in
uncertainty on the frequency and orbital parameters defining our
signal.  When faced with the prospect of searching this space for the
true signal parameters we rely on the concept of the parameter space
metric~\cite{bala96:_gravit_binaries_metric,owen96:_search_templates,1999PhRvD..60b2002O}
which allows us to set a measure of distance by which we can determine
how to sample within the space.  Using this geometrical approach we
are able to satisfy the constraints that our templates will not be
placed too coarsely such we will not ``miss'' a signal and also that
they will not be placed too finely such that we will be wasting
computational effort.  The standard choice is to define the distance
measure for a coherent analysis as the ratio of the expectation value
of the loss in \ac{SNR} local to a signal's true parameters and the
expectation value of \ac{SNR} at the true signal parameters.  This
measure, or mismatch, is then
\begin{equation}\label{eq:mismatch1}
  \mu(\bm{\theta},\bm{\Delta\theta}) = \frac{\rho^{2}(\bm{\theta},\bm{0})-\rho^{2}(\bm{\bm{\theta},\Delta\theta})}{\rho^{2}(\bm{\theta},\bm{0})},
\end{equation}
where we use 
By Taylor-expanding the mismatch around the true signal
location $\bm{\theta}$ we obtain
\begin{equation}
  \mu(\bm{\theta},\bm{\Delta\theta}) = g_{\mu\nu}(\bm{\theta})\,\Delta\theta^{\mu}\Delta\theta^{\nu} +
  o(\Delta\theta^{3}) \label{eq:mismatch2}
\end{equation}
where the coherent metric $g_{\mu\nu}(\bm{\theta})$ is defined as 
\begin{equation}\label{eq:metricdefinition}
  g_{\mu\nu}(\bm{\theta}) = \left\langle\frac{\partial\Phi(\bm{\theta})}{\partial\theta^{\mu}}\frac{\partial\Phi(\bm{\theta})}{\partial\theta^{\nu}}\right\rangle - \left\langle\frac{\partial\Phi(\bm{\theta})}{\partial\theta^{\mu}}\right\rangle\left\langle\frac{\partial\Phi(\bm{\theta})}{\partial\theta^{\nu}}\right\rangle
\end{equation}
with $\langle\ldots\rangle$ representing the average over the coherent observation
time and with $\Delta\theta^{\mu}$ representing the deviation in the
$\mu$'th parameter from its true value.
%
\subsection{The semi-coherent metric\label{sec:semicoherentmetric}}

The first stage of a semi-coherent analysis is the process of performing the
multiple constituent coherent analyses on the $M$ independent data
segments into which the full observation has been
divided\footnote{Whilst in this work we concentrate on a division of
  the complete dataset in the time domain, semi-coherent searches in
  general may also suit division of the dataset in the frequency
  domain, e.g.~\cite{2007CQGra..24..469M}.}.  Such
separate analyses will result in the generation of discretely sampled
values of the detection statistic $\Lambda_{m}$ on the parameter
space spanned by $\bm{\theta}$, where $m$ indexes the segment number.
Note that the metric in each segment and therefore also the mismatch
will vary between segments since the metric can be a function of the
segment epoch (as is the case in our binary system).  

Since $\Lambda_{m}$ are maximized log-likelihood ratios it follows
that a sensible choice is to define the semi-coherent detection
statistic as the sum of these values as a function of $\bm{\theta}$,
\begin{equation}\label{eq:semicoherentdetstat}
  \hat{\Lambda}(\bm{\theta}) = \sum_{m=1}^{M}\Lambda_{m}(\bm{\theta}).
\end{equation}
Therefore, such a statistic would itself represent a log-likelihood on
the space spanned by $\bm{\theta}$, with the understanding that there
has been an implicit maximization over $M$ distinct amplitudes
$A_{m}$ and $\Phi_{0,m}$.  

In practice, the reason behind
adopting an semi-coherent strategy is likely that we lack the computational
resources required to coherently track the phase of a signal over long
observation times.  In which case the introduction and
maximization of $M$ distinct initial phases is an approximation to the
true phase model.  We note that the similar
introduction of the $M$ distinct signal amplitudes also does not appear to be
consistent with our original signal model.  The semi-coherent detection
statistic is now also sensitive to signals for which the amplitude
varies with a time-scale $\gg$ the coherent segment length
$\Delta T$.  This subtle feature is a potential improvement with regards to searches
in the X-ray spectrum for signals from \acp{LMXB} where the amplitude is
not necessarily expected to be constant~\cite{2008AIPC.1068.....W}. 

Based on the fact that the individual segment $\Lambda_{m}$ values are
$\chi^{2}_{2}$ distributed it follows from
Eq.~\ref{eq:semicoherentdetstat} that $\hat{\Lambda}$ is
$\chi^{2}_{2M}$ distributed with a non-centrality parameter equal to
the sum of the individual segment non-centrality parameters.  The
expectation value of $\hat{\Lambda}(\bm{\theta})$ is therefore given
by
\begin{equation}\label{eq:semicoherentexpectation}
  E[\hat{\Lambda}(\bm{\theta},\bm{\Delta\theta})] = 2M +
  \sum_{m=1}^{M}\rho_{m}^{2}(\bm{\theta},\bm{\Delta\theta}).
\end{equation}
The semi-coherent mismatch (defined as the loss in semi-coherently
summed \ac{SNR}) is then 
\begin{subequations}
\begin{eqnarray}
  \hat{\mu}(\bm{\theta},\bm{\Delta\theta})&=&
  \frac{1}{M}\sum_{m=1}^{M}\frac{\rho_{m}^{2}(\bm{\theta},\bm{0})-\rho_{m}^{2}(\bm{\bm{\theta},\Delta\theta})}{\rho_{m}^{2}(\bm{\theta},\bm{0})}, \\
  &=&\frac{1}{M}\sum_{m=1}^{M}\mu_{m}(\bm{\theta},\bm{\Delta\theta}),\label{eq:semicoherentmismatch}
\end{eqnarray}
\end{subequations}
where we have assumed that each segment has identical duration and
noise level.  In practice, whilst it may be simple to maintain constant
segment lengths, detector noise may vary between segments.  In addition, for \ac{GW}
detectors, amplitude modulation of the signal due to the changing
response of the detector as the earth rotates will have an effect~\cite{2007PhRvD..75b3004P}.

By substituting Eq.~\ref{eq:mismatch2} into
Eq.~\ref{eq:semicoherentmismatch} the semi-coherent metric
$G_{\mu\nu}(\bm{\theta})$ can then be expressed as
\begin{equation}\label{eq:semicoherentmetricdef}
 G_{\mu\nu}(\bm{\theta}) = \frac{1}{M}\sum_{m=1}^{M}g_{\mu\nu}^{(m)}(\bm{\theta}).
\end{equation}
This is the standard result, as shown in~\cite{2000PhRvD..61h2001B}, that the semi-coherent metric is
simply the element-by-element average of the constituent coherent metrics. 
%
\subsection{Template placement\label{sec:templateplacement}}

In both the coherent and semi-coherent stages of the analysis the
intention is to use the information contained in the metric
description of the parameter space to place templates.  The task is
then to cover such a space efficiently whilst adhering to the
constraint that no template should have mismatch with any potential
signal, greater than a given threshold, $\mu^{*}$.  In this work we
simply refer the reader to recent efforts made in the field of \ac{GW}
data analysis with regards to metric based template
placement~\footnote{Note that there are many other parameter space
  exploration strategies that do not use the metric
  e.g. Markov-Chain-Monte-Carlo methods.}.

In the case of a constant metric, one in which either careful
parameterization or luck has left a metric of which the elements are
constant over the parameter space, the problem is directly equivalent
to the ``sphere covering'' problem~\cite{CONWAYSLOANE}.  This was
realized and investigated, in the context of \ac{GW} data analysis,
in~\cite{2007CQGra..24..481P}.  In general, in the constant metric
case the optimal solution (in the sense of guaranteed parameter space
coverage) is to use an n-dimensional lattice of templates.  The most
basic (but inefficient) of these lattices being the hyper-cubic
$\text{Z}_{n}$ lattice and the most efficient being the class known as
the $\text{A}_{n}^{*}$ lattice.

For the more general case of a metric whose elements are functions of
the location in the space itself, template placement is more
difficult.  The coordinate volume and orientation of the area local to
a template will change as one moves through the parameter space.
Hence the density and relative separations between templates in each
of the dimensions of the space changes depending on where we are
placing templates.  An optimal solution to this particular problem has
yet to be found although recent work on so-called ``random'' and
``stochastic'' template
banks~\cite{2009PhRvD..79j4017M,harry-2008,babak-2008,2010PhRvD..81b4004M}
shows impressive covering performance especially for higher
dimensional spaces.  These methods use only information regarding the
required local template density, a quantity proportional to the square
root of the determinant of the metric.
%
\subsection{Combining results from different segments\label{sec:combining}}

Let us imagine that the coherent detection statistic has been computed
for all templates in all segments.  We now have the information
required to compute an ensemble of semi-coherent detection statistics
and the general form of the semi-coherent metric
(Eq.~\ref{eq:semicoherentmetricdef}) tells us how to construct a
semi-coherent template bank on the same parameter space.  We note that
the semi-coherent template bank is a) potentially sampled far more finely
than the coherent template banks, and b) potentially defined on a
different parameter space coordinate system (this is the case for the
scenario described in Section~\ref{sec:shortcoherent}).  For each
semi-coherent template we need to perform the following.

Firstly, for each coherent segment we transform the
parameters of the semi-coherent template into the parameters used
in the segment.  For each set of transformed coordinates we 
interpolate the value of the
detection statistic on the coherent
template bank at the desired parameter space location.  The interpolated
detection statistics are then summed according to
Eq.~\ref{eq:semicoherentdetstat}.  This is repeated for each template in
the semi-coherent template bank.

A particularly fast and simple interpolation method is to use the
nearest neighbor approach, the speed and simplicity of which benefits
from the use of a hyper-cubic lattice within each segment.  Higher
precision is achieved by using smaller mismatches in the coherent stage.
Using both the inefficient hyper-cubic lattice and decreasing the
mismatch will of course increase the computational cost of
computing the coherent stage of the analysis.  

The assumption that the coherent analysis had been computed prior to
this procedure need not be the case in practice.  One could perform
the interpolation and summing as each segment is being analyzed as
long as the total time span of the complete data-set was known a
priori, since as we will see, the semi-coherent metric is dependent
upon the total time span. 
%
\section{Results\label{sec:results}}

In this section we report on the results obtained through calculation
of the semi-coherent metric elements via
Eq.~\ref{eq:semicoherentmetricdef} using the binary system phase model
given by Eq.~\ref{eq:loweccentricityphasemodel}.  In parallel we
provide basic schemes for the practical application of a semi-coherent
search in two observational regimes determined by the length of the
coherent observation $\Delta T$.  The first having a ``short''
coherent stage with $\Delta{T}\ll P$ where the segment length is far
shorter than the orbital period, and the second having a ``long''
coherent stage $\Delta{T}\gg P$ where the segment length exceeds the
orbital period.
%
\subsection{The coherent metric for ``short'' ($\Delta{T }\ll P$) coherent segments\label{sec:shortcoherent}}

In practice one need not compute the coherent detection statistic for
all segments using templates given by the semi-coherent metric.
The semi-coherent metric tells us how finely we must sample the parameter
space for the complete semi-coherent analysis.  It is the coherent metric
that tells us how finely we must sample each segment.  

In the ``short'' segment regime the coherent
metric, calculated in our physical frequency and orbital parameter space, develops strong
parameter degeneracies for short observation times, making template
placement difficult.  In this situation the reduction in detection
statistic due to offsets between the template
and the signal in one parameter can be effectively counteracted by
offsets in another.  Additionally, the coherent
metric computed in any chosen coordinates, would in general result in
different \emph{physical} parameter template locations in each
segment.  Therefore, once the coherent segments have been processed,
for a given semi-coherent template there will not be a collection of
corresponding coherent segment detection statistics to sum together
all computed at the exact same parameter space location.

To address the first issue we propose the adoption of a
re-parameterization~\cite{2008PhRvD..78j2005P} that greatly simplifies the phase
model but that is only valid in the limit $\Omega\Delta{T}\ll 1$, i.e. when
the length of coherent segment is only a fraction of the orbital
period.  In this case, by simple Taylor-expansion of the phase model,
given by Eq.~\ref{eq:loweccentricityphasemodel}, in terms of the time $t$ about the midpoint of
each coherent observation we obtain an approximate phase model given by
\begin{equation}\label{eq:uphase}
  \Phi_{m}(t_{j},\bm{u}) = \Phi^{(m)}_{0} + 2\pi\sum_{k=1}^{n} \frac{u^{(m)}_{k}}{k!}\left(t_{j}-t^{(m)}_{\mathrm{mid}}\right)^{k}
\end{equation}
where $t^{(m)}_{\mathrm{mid}}$ is the midpoint of the $m$'th segment,
$\Phi^{(m)}_{0}$ is the phase at this midpoint
and the new coordinates $u^{(m)}_{k}$ map to the physical parameters via
the equations given in Appendix~\ref{app:ucoords}.  Note that the
$\bm{u}$ coordinates themselves are exactly the instantaneous phase
derivatives at the midpoint of each segment indexed by $k$ which runs
from 1 to $n$.  Also note that by this
re-parameterization the boundaries of the parameter space in the new
coordinates will not have the same shape as the physical
parameter boundaries.

\begin{figure}
  \begin{center}
    \includegraphics[bb = 10 40 420 400, width=\columnwidth]{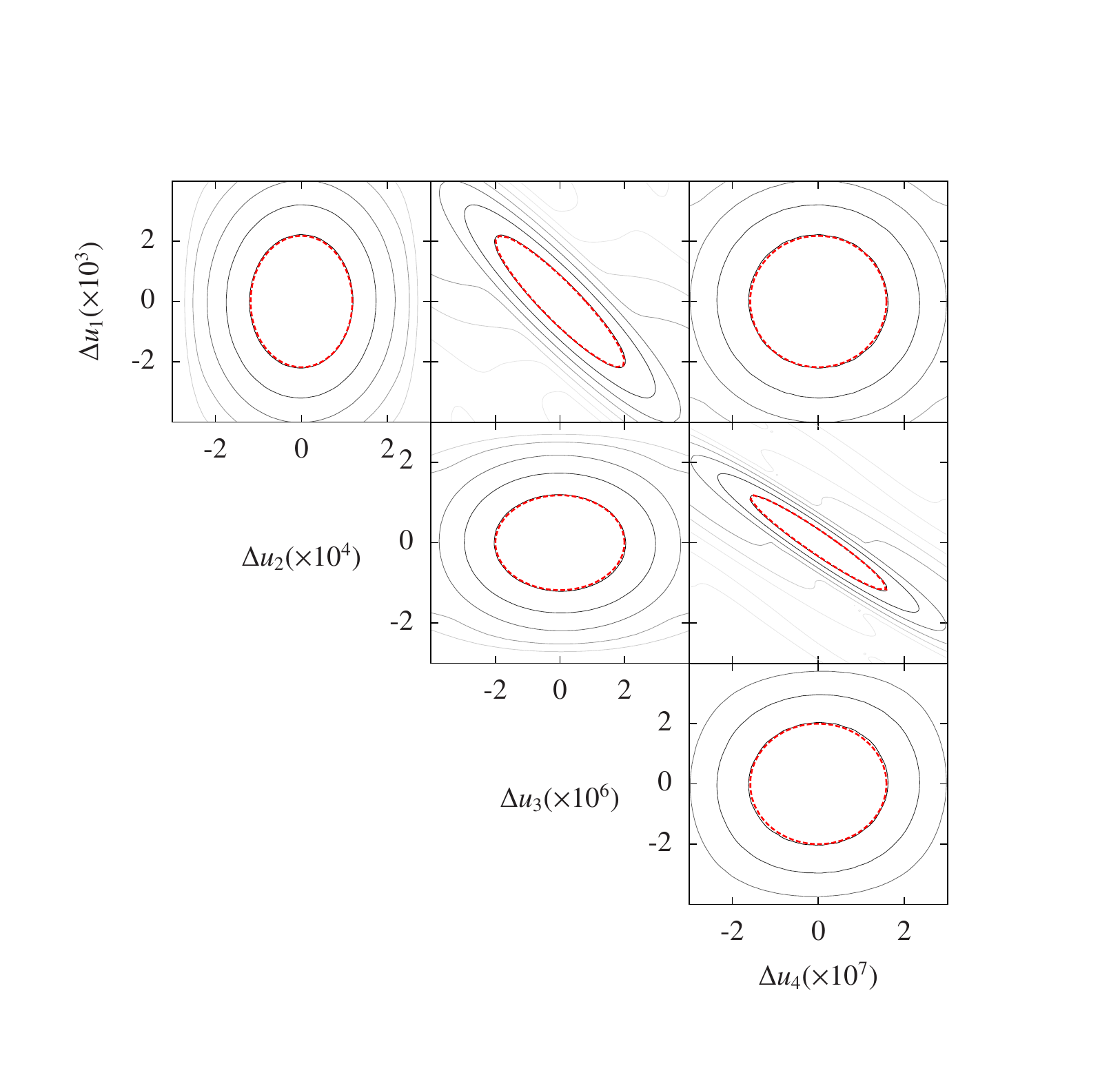}
    \caption{Shown here are the results of a simulation in which the
      mismatch $\mu(\bm{u})$ has been computed over a region in
      the 3-dimensional parameter around the location of a simulated
      signal.  The simulation models a single segment of $200$ seconds
      length with signal parameters
      $\nu=200$ Hz, $a=1$ second, $\Omega=10^{-4}$ rads $\mathrm{s}^{-1}$,
      $e=0.02$ and $\omega=1$ rad ($t_{\mathrm{asc}}$ chosen
      randomly from the range $[-P/2,P/2]$)
      Each panel shows the mismatch as a function of each
      pair of parameters where the mismatch has been maximized over
      all other parameters.  The solid black
      and grey contours indicate the measured mismatch ranging from
      $\mu=0.1$ (black) in steps of $0.1$.  The dashed red ellipses
      are those calculated using the metric given in
      Eq.~\ref{eq:umetric} on the approximate phase given by Eq.~\ref{eq:uphase}. 
    }\label{fig:u_mismatch}
  \end{center}
\end{figure}

By direct application of Eq.~\ref{eq:metricdefinition} to our
re-parameterized phase model we find that the metric in the
$\bm{u}$ coordinates is  
\begin{equation}\label{eq:umetric}
  g_{\mu\nu}(\bm{u})= \left\{ \begin{array}{ll}
      \dfrac{\pi^{2}\mu\nu\Delta
        T^{\mu+\nu}}{2^{\mu+\nu-2}(\mu+1)!(\nu+1)!(\mu+\nu+1)}, & \quad
      \mu+\nu = \text{even}\\
      0, & \quad \mu+\nu = \text{odd}.\end{array}
  \right.
\end{equation}
Here we see that the elements themselves are independent of parameter
space location making this a constant metric and therefore simple for
template placement.  The fact that the $g_{\mu\nu}(\bm{u})$ elements
vanish for $\mu+\nu$ being odd indicates a lack of correlation between
adjacent phase expansion terms, a feature that is exemplified in
Fig.~\ref{fig:u_mismatch}.  In this figure we show the results of a
simulation in which the coherent mismatch $\mu(\bm{u},\bm{\Delta{u}})$
has been computed in a region in the $\bm{u}$ parameter space
surrounding a simulated noise-free signal.  The degree of correlation
between the parameters for the cases where $\mu+\nu$ is even is
clearly indicated by the diagonal orientation of the mismatch
contours.  Also evident from the figure is that the predicted $10\%$
mismatch contour, computed using the metric approximation given in
Eq.~\ref{eq:umetric}, is in good agreement with the simulation
results.  The metric itself is a quadratic approximation around the
peak of the log-likelihood ratio and in general is only expected to be
accurate in the regime $\mu\ll 1$.  As we move away from the true
signal location the simulated contours begin to deviate from their
elliptical shape as higher order contributions to the mismatch become
important.  Fortunately, since by design, the mismatch is the loss in
expected \ac{SNR} one would typically only ever place templates
according to mismatches $\mathcal{O}(10\%)$ so as to recover a large
enough fraction of the \ac{SNR} to avoid missing any signals.  We note
that in this region the predicted mismatch using the metric
approximation would have discrepancies relative to the true mismatch
at a level ${<}0.1\%$.

The re-parameterization of the phase into the $\bm{u}$ coordinates is
in fact entirely equivalent to the approximations made in so-called
acceleration searches~\cite{1990Natur.346...42A,1992PhDT........20A,2004MNRAS.355..147F}.
Here, in just the same way as in an acceleration search, the length of
coherent observation coupled with the orbital period determines how
many orders of expansion are required to accurately model the phase.
Conversely, given a computationally limited number of expansion terms,
there is a corresponding limit to the shortest orbital period that can be searched
for a given coherent observation time.  Based on worst case values of
$t_{\mathrm{asc}}$ in relation to $t_{\mathrm{mid}}$ we find that 
\begin{equation}\label{eq:uexpansionlimit}
  \left(\Omega\Delta{T}\right)\lesssim\left(\frac{\Delta\Phi (n+1)!}{2\pi a\nu}\right)^{1/(n+1)},
\end{equation}
represents a limit on the length of the coherent observations for an $n$'th order
expansion in the $\bm{u}$ coordinates where $\Delta\Phi$ is the
allowed error in phase between signal and model.  For example, in
order to model a $\nu=100$ Hz signal with a projected semi-major axis $a=1$ second with error
in signal phase $\Delta\Phi=\pi/2$ using only an $n=2$ expansion we
would be limited to coherent observations spanning up to ${\approx}1/25$ of an orbit.

The number of templates required to cover a parameter space is proportional to its proper volume, the
integral of the square-root of the metric determinant over the
parameter space, and is given by
\begin{equation}
  \mathcal{N} = \xi(n,\mu) \int_{\mathcal{S}}d\bm{\theta}\,\sqrt{\mathrm{det}\,g}
\end{equation}
where we have used $\mathcal{S}$ to represent the parameter space and the template
density $\xi(n,\mu)$ is a function of the dimensionality of the space
$n$, the desired mismatch $\mu$ and the choice of lattice covering.
This gives us
\begin{equation}\label{eq:numberofutemplates}
  \mathcal{N}^{(\bm{u})}_{\mathrm{co}}\propto
  \xi(n,\mu)\Delta{T}^{n(n+1)/2}\int_{\mathcal{S}}d^{n}u
\end{equation}
as the number of templates required for a single segment.  Note that
we do not give the explicit scaling associated with the parameter
space boundaries (as done below in
Eqs.~\ref{eq:numberofsemitemplates},\ref{eq:numberofsemitemplatescirc}).  The
integral over the parameter space volume is a complicated function of
these boundaries and will behave differently depending upon the
relative sizes of the uncertainties in the orbital parameters
e.g. well known orbital period with poorly known time of ascension
compared to poorly known period and well known time of ascension.
One of the simpler cases one can consider is one in which the time of
ascension is completely unknown.  In this case the parameter space
boundaries in the $\bm{u}$ space form a hypercube with limits
$\nu_{\text{min}}<u_{1}<\nu_{\text{max}}$ and
$-a_{\text{max}}\nu_{\text{max}}\Omega_{\text{max}}^{k}<u_{k}<a_{\text{max}}\nu_{\text{max}}\Omega_{\text{max}}^{k}$
for $k>1$ and the number of templates is given by
\begin{equation}\label{eq:numberofutemplates_specialcase}
  \mathcal{N}^{(\bm{u})}_{\mathrm{co}}\propto
  \xi(n,\mu)\Delta{T}^{n(n+1)/2}\left(\nu_{\text{max}}-\nu_{\text{min}}\right)\left(a_{\text{max}}\nu_{\text{max}}\right)^{n-1}\Omega_{\text{max}}^{n(n+1)/2-1}.
\end{equation}
Here we would like to highlight a general feature of the
re-parameterization process.  In simplifying the template placement
procedure, in the sense of using a constant metric, we have introduced
parameter space boundaries in the $\bm{u}$ space that both vary
between segments and are in general no longer simple hyper-cubic
spaces. 

\subsection{The semi-coherent metric for short ($\Delta{T}\ll P$) coherent segments\label{sec:shortsemicoherent}}

In order to compute the semi-coherent metric in this case we make the
following assumptions and choices.  We assume that we have many
individual ``short'' coherent observations of duration $\Delta{T}$ randomly and uniformly distributed over
the total observation span $\tau$, and that $\tau\gg P$ such that the
total span of all observations contains many source orbits.  We also
assume that the time of passage through the ascending node of the
orbit $t_{\mathrm{asc}}$ has been chosen such that it is close to
$t=0$ (defined as the midpoint of the span of all observations).  This
is always possible since $t_{\mathrm{asc}}$ can be redefined for any
orbital cycle by adding or subtracting integer multiples of the
orbital period~\footnote{One must be careful to propagate the
  uncertainty in the orbital period through to any redefinition of
  $t_{\mathrm{asc}}$.}.  

The resulting semi-coherent metric elements are given in
Appendix~\ref{app:metricelements} for the case of $e\ll 1$, $\tau\gg
\Delta{T}$ and $\tau\gg P$.  We also give the approximate
Taylor-expanded versions of the elements applicable in the limit that
$\Omega \Delta{T}\ll 1$ where the segment lengths are far smaller than
the orbital period.  It is on these Taylor-expanded results that we
now focus on.  Note that throughout, Taylor expanded metric elements
are labeled with the superscript $(\text{T})$.

The specific parameter vector for which our metric has been calculated
is
$\bm{\theta}=\left\{\nu,a,t_{\mathrm{asc}},\Omega,\kappa,\eta\right\}$
and in order to compute the semi-coherent metric itself we use the
following approximation
\begin{eqnarray}
  \lim\limits_{\Delta{T}\ll P}  G_{\mu\nu}(\bm{\theta}) &\approx &
  \frac{1}{M}\sum_{k=1}^{M}g_{\mu\nu}^{(T)}(\bm{\theta}) \nonumber \\
  &\approx &\lim\limits_{\tau\rightarrow\infty} \frac{1}{\tau}\int_{-\tau/2}^{\tau/2}g^{(T)}_{\mu\nu}(\bm{\theta},t_{\mathrm{mid}})\,dt_{\mathrm{mid}}
\end{eqnarray}
where we have replaced the sum over segments with an integral over the
midpoint of each segment (approximated as a continuous variable in the
regime where $M\gg 1$).  Under this approximation the semi-coherent
metric can be accurately approximated as
\begin{widetext}
\begin{eqnarray}\label{eq:semi-coherentmetric}
  \lim\limits_{\Delta{T}\ll P}  G(\bm{\theta}) \approx
  \left[ {\begin{array}{cccccc}
        G^{(\mathrm{T})}_{\nu\nu} & & & & &\\
         & G^{(\mathrm{T})}_{aa} & & & 0 & \\
         & & G^{(\mathrm{T})}_{t_{\mathrm{asc}}t_{\mathrm{asc}}} & & &\\
         & & & G^{(\mathrm{T})}_{\Omega\Omega} & &\\
         & 0 & & & G^{(\mathrm{T})}_{\kappa\kappa} &\\
        & & & & & G^{(\mathrm{T})}_{\eta\eta} \\
      \end{array} } \right] =
  \frac{\pi^{2}}{6}\Delta{T}^{2}\left[ {\begin{array}{cccccc}
        2 & & & & &\\
         & (\nu\Omega)^2 & & & 0 & \\
         & & (\nu{a}\Omega^{2})^2 & & &\\
         & & & (\nu{a}\Omega\tau)^{2}/12 & &\\
         & 0 & & & (\nu{a}\Omega)^{2} &\\
        & & & & & (\nu{a}\Omega)^2 \\
      \end{array} } \right]\nonumber \\
\end{eqnarray}
\end{widetext}
Note that we have made the approximation that all off-diagonal
elements can be set to zero, indicating that there are no parameter
correlations.  This approximation can be validated by inspecting each
element pair in turn and noting that the off-diagonal correlation
terms are, in all cases, negligible.  A quantitative validation of
this statement can be performed by computing the volume element (the
square-root of the metric) on each 2-dimensional subspace and noting
that the inclusion of the off-diagonal terms has insignificant
effect.  Significant correlations would reduce the volume element and
ultimately lead to a reduction in the number of templates because
individual templates would occupy more space.  If we consider, for
example, the parameters $\nu$ and $a$ we find that by including the
off-diagonal term in the calculation of the sub-volume element the
fractional reduction in the volume is ${\approx}a^2\Omega^{2}/4$
which is negligible for all practical orbital parameter values.  

The reason that the metric becomes approximately diagonal and that
therefore parameter correlations are removed can be explained as
follows.  The analysis of each segment can be viewed as an independent
measurement of the signal and its parameters.  Recall that we use
templates placed in the $\bm{u}$ space rather than the physical
parameter space $\bm{\theta}$ because of degeneracies in the physical
parameters for short observation times.  This simply means that our
ability to determine certain physical parameters and certain physical
parameter combinations is poor in this choice of coordinates.

For example, imagine a segment coinciding with a small fraction of the
orbit that places the source at the ascending node of the orbit. In
this case the source is moving towards us at its greatest speed.  We
will also focus on a single parameter pair, the intrinsic frequency
$\nu$ and the projected orbital semi-major axis $a$ but note that the
proceeding arguments apply equally well to all other parameters.  At
this segment epoch we are essentially measuring the maximally Doppler
modulated signal frequency ${\approx}\nu(1 + a\Omega)$ Hz and we
therefore find that there exists a degeneracy making us unable to
disentangle the true value of the frequency from that of the projected
semi-major axis.  The signal can equally well be estimated within this
segment as having a higher frequency and smaller semi-major axis or
vice versa.  In this sense we would say that there exists a negative
correlation between these parameters.  A negative (or positive)
correlation implies that an increase in mismatch from a positive
offset in one of the parameters can be counteracted by a negative
(positive) offset in the other.

If we now examine the behavior of the $\nu,a$ correlations at an
orbital epoch advanced in time by $P/4$ seconds we find that the
source is now moving perpendicular to the observer's line of sight.
In this case the frequency is approximately constant for the duration
of the segment and there would be little information in the signal
regarding it's orbital acceleration.  There now exists no correlation
between $\nu$ and $a$ and one is able to determine the frequency quite
accurately at the expense of relatively poor determination of the
projected semi-major axis.  Advance again by $P/4$ and the source is
now receding from the observer at its maximum speed.  We now see
exactly opposite behavior to that seen for the initial epoch in that
there now exists an equal in magnitude positive correlation between
$\nu$ and $a$.  Finally, advancing again by $P/4$ we find that the
correlations have vanished just as for the epoch one half orbit
earlier.

Under our assumptions that the total observation spans many orbits and
that the short observations themselves are uniformly distributed over
this span we obtain measurements from all orbital phases.  The
combination of a pair of measurements that have approximately equal
and opposite correlations will therefore remove those correlations.
With many segments randomly positioned in orbital phase each one can
be approximated as having a corresponding segment on the opposite side
of the orbit.  As such, each opposing pair of segments removes the
correlations in the final semi-coherent measurement.

\begin{figure*}
  \begin{center}
    \includegraphics[bb = -20 20 700 600, width=\textwidth]{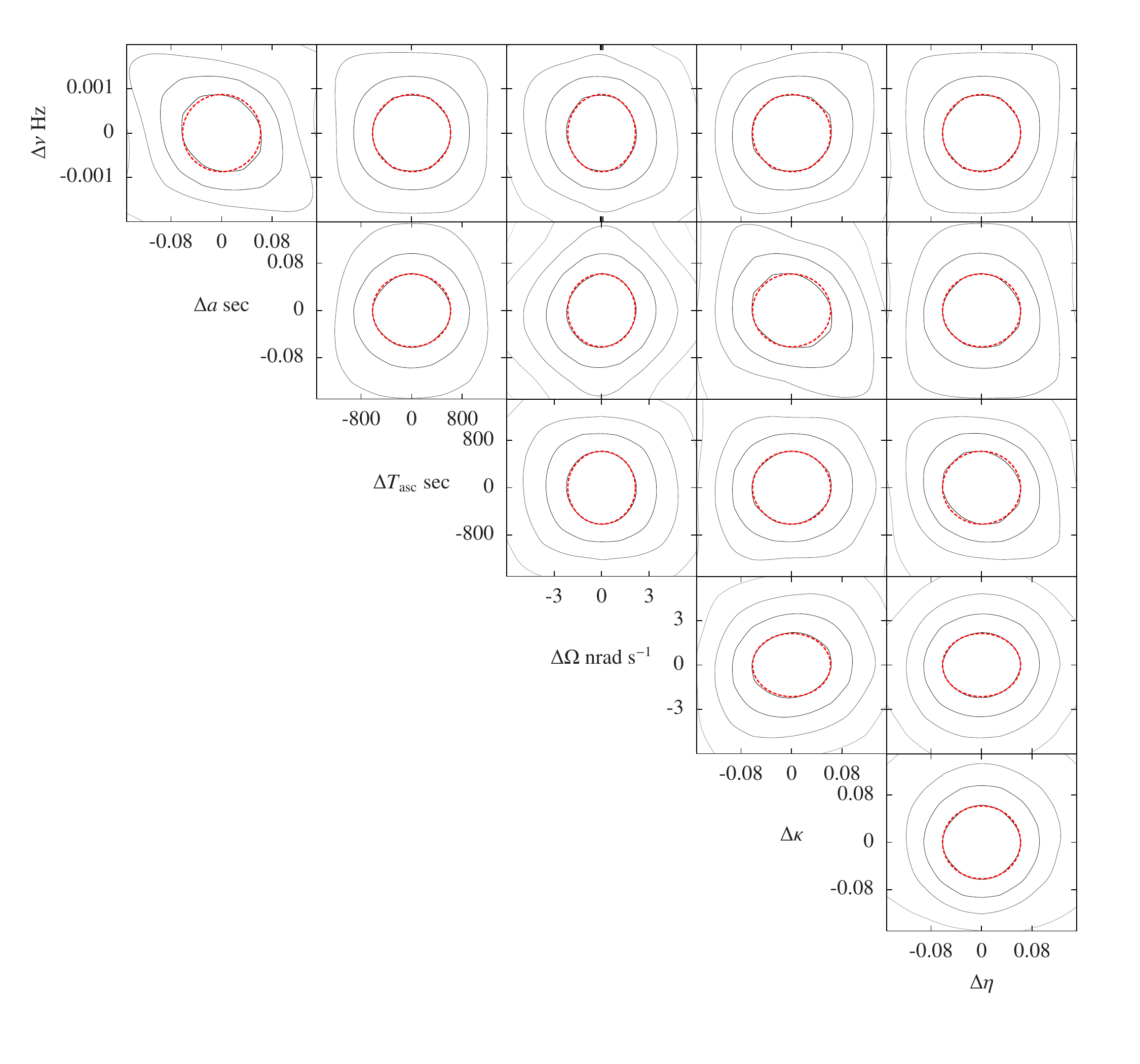}
    \caption{The results of a simulation in which the
      mismatch $\mu(\bm{\theta},\bm{\Delta\theta})$ has been computed over a region in
      the 6-dimensional parameter space around the location of a simulated
      signal.  The simulation modeled $500$ individual coherently
      analyzed segments each of $200$ seconds length spread over a
      period of $\tau=10^{8}$ seconds.  The signal parameters were
      $\nu=200$, $a=1$, $\Omega=10^{-4}$ rads $\mathrm{s}^{-1}$,
      $e=0.02$ and $\omega=1$  ($t_{\mathrm{asc}}$ chosen
      randomly from the range $[-P/2,P/2]$).  Each panel shows the mismatch as a function of each
      pair of parameters where the mismatch has been maximized over
      all other parameters.  The solid black
      and grey contours indicate the measured mismatch ranging from
      $\mu=0.1$ (black) in steps of $0.1$.  The dashed red ellipses
      are those calculated using the approximate metric given in
      Eq.~\ref{eq:semi-coherentmetric}.
    }\label{fig:semicoherent_mismatch}
  \end{center}
\end{figure*}

Shown in Fig.~\ref{fig:semicoherent_mismatch} are the results of
numerical simulations whereby the mismatch
$\hat{\mu}(\bm{\theta},\bm{\Delta\theta})$ has been computed over the
6-dimensional parameter space surrounding a simulated signal.
Overlayed on the plots within this figure are the analytically derived
mismatch ellipses corresponding to solving
Eq.~\ref{eq:semicoherentmetricdef} for $\hat{\mu} = 0.1$ using
Eq.~\ref{eq:semi-coherentmetric} as the semi-coherent metric.  The
simulation comprised of $M = 500$ coherently analyses $\Delta{T} =
200$ second long segments randomly distributed across an full
observation span $\tau = 10^{8}$ seconds.  The simulated signal has an
orbital angular frequency $\Omega = 10^{-4}$ rads $\mathrm{s}^{-1}$
(orbital period of $17.45$ hours), a frequency $\nu=200$ Hz, an
orbital semi-major axis $a=1$ second and an eccentricity $e=0.02$ and
argument of periapse $\omega = 1$.  It is clear that in this case the
approximations made in the Taylor expansion and discarding of the
off-diagonal metric elements were valid since there is no indication
of correlation between parameters.  Correlations would appear as
ellipsoidal mismatch contours with semi-minor and semi-major axes
inclined with respect to the parameter axes, i.e. tilted ellipses as
seen in some panels of Fig.~\ref{fig:u_mismatch}.  We also see again
at large $\mu> 0.1$ mismatches, effects due to the breakdown of the
quadratic approximation used in the definition of the metric.  These
deviations from ellipticity, seen the simulated mismatch contours, can
also be attributed to the finite number ($M=500$) of segments randomly
distributed over the observation span.  This is due to the fact that
the mechanism through which one can assume the disappearance of the
off-diagonal metric elements is applicable only in the limit where the
number of segments $M\gg 1$.  We note however, that in the region of
interest around $\mu\sim 10\%$ the discrepancy between
the metric approximation and our simulation is ${<}1\%$.  To place
this in a practical context, using this metric to place templates with
an expected maximum mismatch between signal and template of $10\%$, in
a worst case scenario could result in a true mismatch of
${\approx}11\%$.

Using the metric result we can compute the number of semi-coherent
templates required for a search and the scalings associated with the
parameter space dimensions.  Considering the full $n=6$
low-eccentricity orbit system we obtain
\begin{eqnarray}
  \mathcal{N}_{\mathrm{semi}} &=&
  \frac{\pi^{6}}{16\sqrt{6}\mu^{3}}\Delta{T}^{6}\tau\left(\nu^{6}_{\mathrm{max}}-\nu^{6}_{\mathrm{min}}\right)\left(a^{5}_{\mathrm{max}}-a^{5}_{\mathrm{min}}\right)\nonumber
  \\
  &&\times\left(\Omega^{7}_{\mathrm{max}}-\Omega^{7}_{\mathrm{min}}\right)\left(t_{\mathrm{asc,max}}-t_{\mathrm{asc,min}}\right) e_{\mathrm{max}},\label{eq:numberofsemitemplates}
\end{eqnarray}
where we have defined the parameter space $\mathcal{S}$ as a
6-dimensional cube bounded by the minimum and maximum ranges of the
parameter vector $\bm{\theta}$.  We have used the relation $d\kappa
d\eta = e|\cos(2\omega)|de d\omega$ in computing the integral over the
eccentricity parameters.   We have used lower and upper bounds of zero and
$e_{\text{max}}$ respectively for eccentricity and assumed that the
argument of periapse would always be completely unknown and hence
would have a range of zero to $2\pi$.  Note that for the template
density we have assumed sub-optimal template placement based on a
hyper-cubic lattice.  A reduction of a factor of $\approx 6.8$ can be
achieved if an $\mathrm{A}_{n}^{*}$ lattice is used in this $n=6$ case.

\begin{figure}
  \begin{center}
    \includegraphics[bb = 0 0 360 250,width=\columnwidth]{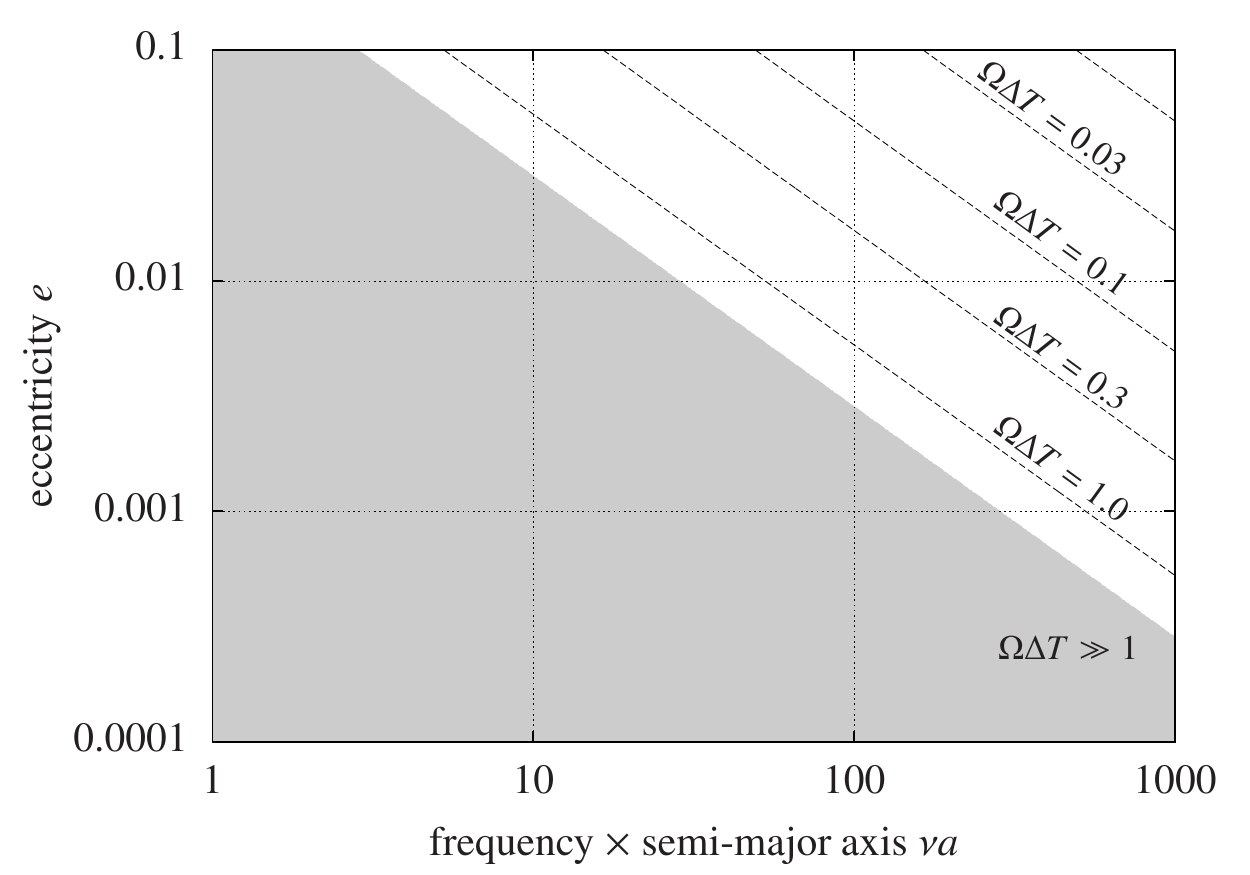}
    \caption{An orbital parameter sub-space represented as the
      eccentricity versus the product of the signal frequency and
      projected semi-major axis.  The dashed
      lines correspond to the semi-coherent metric's requirement for
      templates in the eccentricity parameters $\kappa$ and $\eta$.  Each
      line represents a different coherent observation length measured
      in units of $\Omega \Delta T$.  Above each dashed curve a semi-coherent
      analysis would require templates in the eccentricity parameters
      for a mismatch $\mu=0.1$.  Correspondingly, as the coherent
      integration time increases (relative to the orbital period) a
      limit is achieved at the boundary of the shaded region.  Below
      this limit one need not include eccentricity in the signal model.
    }\label{fig:eccentricity_region}
  \end{center}
\end{figure}

We show in Fig.~\ref{fig:eccentricity_region} the regions of parameter
space for which a semi-coherent search would require coverage in the
eccentricity parameters $\kappa$ and $\eta$.  This requirement is strongly dependent on the length of the
constituent coherent observations, the product of the frequency and
semi-major axis, and the eccentricity itself.  As stated earlier, for the majority of our
primary sources, the \acp{LMXB}, the expected eccentricities in these
systems are relatively low (typically ${<}10^{-3}$).  In addition, the
computational limitations imposed by the size of the remaining
parameter space forces the coherent observation length to shorter times.  This
leads us to the conclusion that in general a circular orbit phase
model will be applicable to the \ac{LMXB} sources.  For a
circular orbit system, ignoring the $\kappa$ and $\eta$ dimensions, we
have the following template scalings 
\begin{eqnarray}
  \mathcal{N}^{\mathrm{circ}}_{\mathrm{semi}} &=&
  \frac{\pi^{4}}{36\sqrt{6}\mu^{2}}\Delta{T}^{4}\tau\left(\nu^{4}_{\mathrm{max}}-\nu^{4}_{\mathrm{min}}\right)\left(a^{3}_{\mathrm{max}}-a^{3}_{\mathrm{min}}\right)\nonumber
  \\
  &&\times\left(\Omega^{5}_{\mathrm{max}}-\Omega^{5}_{\mathrm{min}}\right)\left(t_{\mathrm{asc,max}}-t_{\mathrm{asc,min}}\right).\label{eq:numberofsemitemplatescirc}
\end{eqnarray}
Note that again we have assumed a sub-optimal hyper-cubic template
lattice and in this $n=4$ case a factor of $\approx 2.8$ reduction in
templates can be achieved using an $\text{A}_{n}^{*}$ lattice.  We
wish to stress that common mistakes made in estimating the required
number of templates stem from the inclusion of dimensions that are
geometrically ``thinner'' than the mismatch coverage of a single
template.  Such mistakes can lead to severe underestimation of the
number of required templates.  For example, take the parameters of the
simulations for which results are shown in
Fig.~\ref{fig:semicoherent_mismatch} and apply them to
Fig.~\ref{fig:eccentricity_region} which shows the requirements of a
semi-coherent search with regards to placing templates in the
eccentricity parameters .  We see that for $\Omega\Delta{T}=0.02$ and
$\nu{a} = 200$ the corresponding eccentricity threshold is $\approx
0.1$.  We would therefore expect a semi-coherent search t be unable to
distinguish eccentric orbits from circular orbits for eccentricities
below this value.  Looking at the mismatch ellipses shown in
Fig.~\ref{fig:semicoherent_mismatch} we see that this is indeed the
case, the scale of the ellipses in the $\kappa$ and $\eta$ dimensions
($\kappa^2+\eta^2=e^2$) is $\sim 0.1$ whereas the signal has
eccentricity $e=0.02$.  This is a clear example of when one should
assume a circular orbit model and remove the eccentricity and argument
of periapse from the analysis.

The linear dependence of the number of templates on the total
observation span $\tau$ in both the eccentric and circular obit cases
indicates that an increase in observation span results in an increase
in the number of templates.  One can view this as a ``refinement'' of
the semi-coherent template bank with increasing observation span.  For
our case this refinement comes from a single metric element,
$G_{\Omega\Omega}$, and governs the template spacing in orbital angular
frequency alone.  The reason that we see this behavior is that the
way in which we combine segments in this regime is a function of 
the orbital angular frequency (or equivalently the orbital period).
In order to combine short coherent segments over increasingly large
separations in time, one must have an increasingly accurate
determination of the orbital phase.  In short, the longer the
observation span the more accurately one can determine the orbital
period and hence more templates must be used to cover the
corresponding search dimension $\Omega$.

%
\subsection{The coherent and semi-coherent metric for long ($\Delta T \gg P$) coherent segments\label{sec:longcoherent}}

If the uncertainties in each of the parameter space dimensions is
sufficiently small or the available computational power is large
enough then one may be able to work in the regime $\Omega \Delta
T\gg 1$.  In this case one would be able to construct a search for which it is possible to
achieve a coherent observation length greater than the length of the
orbital period.  The metric components can then be approximated by taking the
limit in the regime $\Omega \Delta T\gg 1$ (given in
Appendix~\ref{app:metricelements}).  Note that in this regime
the physical coordinates
$\bm{\theta}=\left\{\nu,a,t_{\mathrm{asc}},\Omega,\kappa,\eta\right\}$
(as used for the semi-coherent metric) are a sensible choice upon
which to base our metric.  The metric on a
coherent data segment can then be well approximated by
\begin{widetext}
\begin{equation}\label{eq:approxcoherentmetric}
  \lim\limits_{\Delta{T}\gg P}  g(\bm{\theta}) \approx
  \left[ {\begin{array}{cccccc}
        g_{\nu\nu} & & & & &\\
        & g_{aa} & & & 0 & \\
        & & g_{t_{\mathrm{asc}}t_{\mathrm{asc}}} & & &\\
        & & & g_{\Omega\Omega} & &\\
        & 0 & & & g_{\kappa\kappa} &\\
        & & & & & g_{\eta\eta} \\
      \end{array} } \right] =
  \frac{1}{6}\pi^{2}\left[ {\begin{array}{cccccc}
        2\Delta{T}^{2} & & & & &\\
         & 12\nu^{2} & & & 0 & \\
         & & 12(\nu{a}\Omega)^2 & & &\\
         & & & (\nu{a}\Delta{T})^{2} & &\\
         & 0 & & & 3(\nu{a})^{2} &\\
        & & & & & 3(\nu{a})^2 \\
    \end{array} } \right].
\end{equation}
\end{widetext}
where non-zero off-diagonal terms have been approximated as equal to
zero.  We justify this in the same way as done for our similar
approximation to the semi-coherent metric in
Eq.~\ref{eq:semi-coherentmetric}.  Just as in the semi-coherent case,
the off-diagonal terms constitute negligible correlations and their
inclusion leaves the metric determinant essentially unchanged.

\begin{figure*}
  \begin{center}
    \includegraphics[bb = -20 20 700 600, width=\textwidth]{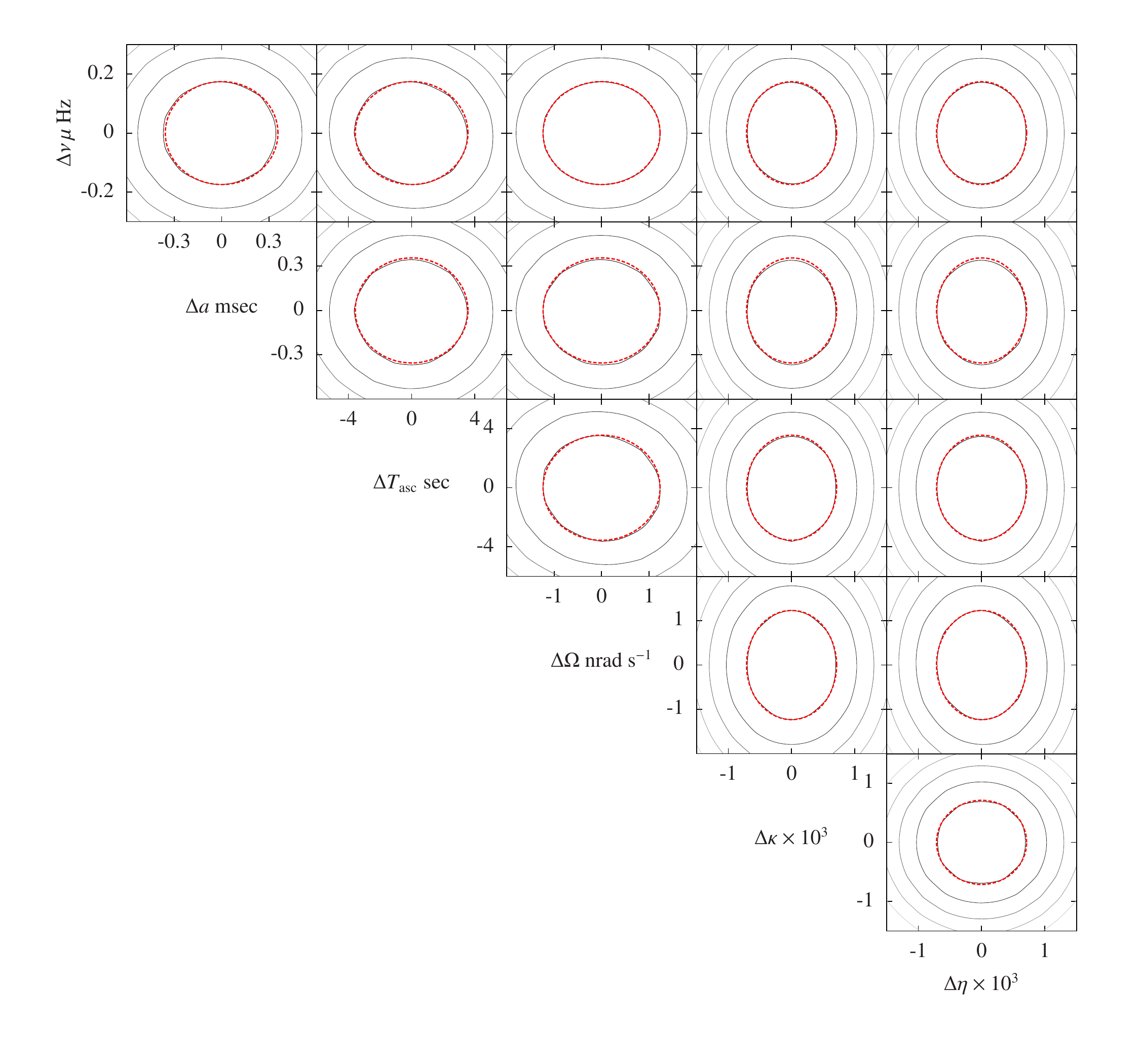}
    \caption{The results of a simulation in which the
      mismatch $\mu(\bm{\lambda})$ has been computed over a region in
      the 6-dimensional parameter around the location of a simulated
      signal.  The simulation modeled a single coherently
      analyzed segment of $10^6$ seconds length containing a signal
      with parameters
      $\nu=200$ Hz, $a=1$ sea, $\Omega=10^{-4}$ rads $\mathrm{sec}^{-1}$,
      sea, $e=0.005$ and $\omega=1$ rad ($t_{\mathrm{asc}}$ chosen
      randomly from the range $[-P/2,P/2]$).  Each panel shows the mismatch as a function of each
      pair of parameters where the mismatch has been maximized over
      all other parameters (these are not slices).  The solid black
      and grey contours indicate the measured mismatch ranging from
      $\mu=0.1$ (black) in steps of $0.1$.  The dashed red ellipses
      are those calculated using the approximate metric given in
      Eq.~\ref{eq:approxcoherentmetric}.
    }\label{fig:coherent_mismatch}
  \end{center}
\end{figure*}

In Fig.~\ref{fig:coherent_mismatch} we see the degree of agreement between mismatches predicted by the
approximate metric and those generated by signal injections (using
Eq.~\ref{eq:exactphasemodel} as the signal model).  Again we see that
our diagonal metric approximation is validated by the clear lack of
correlation between parameters and that the discrepancy between
approximated and simulated mismatches in this case is ${<}0.1\%$. 
Note that unlike the metric on the $\bm{u}$ parameters this metric is
independent of the epoch of each segment and hence a template bank
generated from Eq.~\ref{eq:approxcoherentmetric} can always be
identical in equal length segments.  This type of implementation, as
we will see in the following section, can make the combination of
results from many segments a far easier task than for a search using the
$\bm{u}$ coordinates.  In addition, it makes the construction of the
semi-coherent metric very simple.  Since the semi-coherent metric is
equal to the ``average'' metric (averaged over segments) and in this
case the coherent metric remains unchanged for equal length segments,
the semi-coherent metric is 
\begin{equation}
  \lim\limits_{\Delta{T}\gg P}  G(\bm{\theta}) = \lim\limits_{\Delta{T}\gg P}  g(\bm{\theta}) 
\end{equation}
and is therefore identical to the coherent metric given in
Eq.~\ref{eq:approxcoherentmetric}.

We can again compute the scalings associated with the number of
templates required to cover our 6-dimensional parameter space as done
in Eqs.~\ref{eq:numberofsemitemplates} and \ref{eq:numberofutemplates}
giving us
\begin{eqnarray}\label{eq:numberofcotemplates}
  \mathcal{N}_{\mathrm{co}} = \mathcal{N}_{\mathrm{semi}} &=&
  \frac{9\sqrt{2}}{\pi^{6}\mu^{3}}\Delta{T}^{2} \left(\nu^{6}_{\mathrm{max}}-\nu^{6}_{\mathrm{min}}\right)\left(a^{5}_{\mathrm{max}}-a^{5}_{\mathrm{min}}\right)\nonumber
  \\
  &&\times\left(\Omega^{2}_{\mathrm{max}}-\Omega^{2}_{\mathrm{min}}\right)\left(t_{\mathrm{asc,max}}-t_{\mathrm{asc,min}}\right) e_{\mathrm{max}},\nonumber\\
\end{eqnarray}
where we note that as done in the previous template number
estimates, we assume that each dimension has a thickness greater than
the width of a single template span in the corresponding dimension.
For the circular orbit case we obtain
\begin{eqnarray}\label{eq:numberofcotemplatescirc}
  \mathcal{N}^{\mathrm{circ}}_{\mathrm{co}} = \mathcal{N}^{\mathrm{circ}}_{\mathrm{semi}} &=&
  \frac{\pi^{4}}{27\mu^{2}}\Delta{T}^{2}\left(\nu^{4}_{\mathrm{max}}-\nu^{4}_{\mathrm{min}}\right)\left(a^{3}_{\mathrm{max}}-a^{3}_{\mathrm{min}}\right)\nonumber
  \\
  &&\times\left(\Omega^{2}_{\mathrm{max}}-\Omega^{2}_{\mathrm{min}}\right)\left(t_{\mathrm{asc,max}}-t_{\mathrm{asc,min}}\right).
\end{eqnarray}
In both eccentric and circular orbit cases we have assumed a
sub-optimal hyper-cubic lattice of templates.  As stated in the
previous section, reductions of factors of ${\approx}6.8$ and
${\approx}2.8$ can be achieved in the numbers of templates for the
eccentric and circular orbit cases respectively if an
$\text{A}_{n}^{*}$ lattice is used.

We note that the number of templates in both the eccentric and
circular orbit cases does not depend on the total observation span
$\tau$.   This means that there is no ``refinement'' in the template
bank as one uses larger numbers of segments.  In contrast to the
previously investigated regime $\Delta T\ll P$, if there
is an offset in the orbital period then the mismatch from
one segment will be identical to the mismatch from all other
segments which will consequently be identical to the total mismatch.
In short, this is because there are no parameters in our model for which fractional losses in \ac{SNR}
accumulate beyond a single orbital period e.g. if there are offsets in
all parameters, the fractional \ac{SNR} loss after one orbit will be identical to
the total fractional \ac{SNR} loss after the next.  This is in
contrast to the behavior of the metric and mismatch in the $\Delta
T\ll P$ regime discussed in Section~\ref{sec:shortcoherent}.  However,
for the same volume of parameter space, a significantly larger number
of templates would be required in the coherent stage for segment
lengths that far exceed the orbital period compared to the case when
the orbital period far exceeds the segment length.

\section{Discussion\label{sec:discussion}}

%
We have presented a brief review of the phase model for continuously
emitting sources in binary systems together with a review of the
parameter space metric defined on the fractional loss in \ac{SNR} when
offset from the true signal location in parameter space.
We then proceeded to apply the latter to the former to construct the
semi-coherent parameter space metric for application to searches for such
systems in noisy data.  In addition we provide the outline for the
practical application of a search strategy employing the semi-coherent
metric in the opposing limits that the segments that are analyzed coherently
span small fractions of the source orbital period and have durations
far greater than the orbital period.   

%
%
We have shown that in the short duration segment regime the
semi-coherent metric on the physical parameters $\bm{\theta}$ becomes
approximately diagonal and hence the correlations between parameters
are removed.  This we attribute to the averaging procedure applied to
the metric elements and the specific behavior of these elements as a
function of orbital phase corresponding to each segment epoch.  Whilst
the metric itself is still a function of location within the parameter
space making lattice based template placement difficult, the lack of
parameter correlations does somewhat simplify the procedure.
We note that for the short duration coherent segment scenario there is
only one term in the semi-coherent metric that depends on the
semi-coherent observation span.  This is the element that determines
the precision with which the orbital angular frequency must be sampled
(and conversely the precision with which it can be determined in the
limit of low signal-to-noise ratio).  For all other parameters the
equivalent precisions depend only upon the coherent observation time,
the signal frequency, orbital angular frequency and semi-major axis.
This result is identical in nature to that obtained
in~\cite{2009PhRvL.103r1102P} where it was found that only one
parameter in their isolated pulsar model, the pulsar spin derivative,
was responsible for refinement of the template bank.  The explanation
is similar for both cases, both the spin derivative and orbital period,
in each search respectively, determine how results from individual
segments are combined whereas the other parameters do not. 
It is only the parameter $\Omega$ that is capable of causing an ever
increasing offset in the orbital phase (just as the spin derivative
in~\cite{2009PhRvL.103r1102P} is
capable of causing a similarly increasing offset in signal phase).  As
such, coherent templates within increasingly separated segments would
experience increasing mismatch for a constant offset in $\Omega$.  In
contrast, an offset in the time of orbital ascension (assuming
an exactly known orbital period) would cause the orbital phase to be
offset by a constant amount in all segments.  Consequently the
mismatch would also be constant and hence the average mismatch,
equivalent to the total semi-coherent mismatch, would also be constant
with respect to the observation span $\tau$.  The consequence of this
refinement on the orbital angular frequency parameter is that the cost
of the semi-coherent stage of the analysis can dominate the cost of
the coherent stage.  However, one is always able to tune the relative
costs by choosing a shorter or longer segment length.  We leave the
analysis of how to optimally make such choices for future work.

%
%
When the duration of the coherent segments are much greater than the
orbital period there is no dependence upon the total observation span
and the semi-coherent metric is simply equal to the coherent metric
since it is constant for each segment.  In addition, just as for the
short segment regime we find that the metric is diagonal and that
there are therefore no parameter correlations. The independence of the
total observation span on the metric can be explained by the fact that
there is no dependence upon any of the parameters in the process of
combining the results from each segment.  In contrast to the short
segment regime, one need not know the orbital angular frequency to map
a set of physical parameters in one segment to the corresponding
(identical) set in another segment.  This does not mean that parameter
estimation is not enhanced with the addition of more segments over a
longer observation span.  In this case it is the semi-coherent
accumulation of \ac{SNR} that improves parameter estimation whilst the
template density remains constant.

%
One interesting point to make given this result is the comparison
between the different template scalings with the observation
parameters for the short and long coherent duration regimes.  For a
fixed set of boundaries in the full 6-dimensional parameter space
there appears to exist an observation span at which it becomes
favorable (in terms of using fewer semi-coherent templates) to expand
the coherent observation time to be consistent with the long coherent
regime.  This is because the template scaling with $\Delta T$ is
steeper for the short segment compared to the long segment scenario.
The point at which this occurs is dependent upon many interdependent
factors including the volume of the parameter space, the relative
uncertainties in each dimension and how many orders are included in
the Taylor expansion in the $\bm{u}$ coordinates.  This statement is
also under the assumption that the majority of the computational cost
goes into computing the semi-coherent detection statistic and that the
cost of computation of the individual segment statistics is negligible
and that the computational cost of a long duration template is equal
to that of a short duration template .  It also neglects the issue
that for a short coherent duration some dimensions are likely to not
require templates.  This occurs when the template span is greater than
the parameter space range in that dimension.  Note for example the
metric elements corresponding to the eccentricity parameters $\kappa$
and $\eta$ for the short and long segment regimes.  The template
spacing in a given dimension is proportional to the inverse of the
square-root of the corresponding metric element.  The ratio between
the short and long segment $\kappa$ template spacing is
$\sqrt{3}/(\Omega\Delta T_{\text{s}})$ where $\Delta T_{\text{s}}$
here is the short segment duration.  Dependent upon the uncertainty in
$\kappa$, this may mean that templates are simply not required in the
$\kappa$ dimension for the short segment but are required in the long
segment regime.  If we look at all the contrasting template sizes in
the short and long duration coherent segments regimes we see that in
all dimensions except the spin frequency $\nu$ and the orbital angular
frequency $\Omega$, a template span is greater by a factor $\propto
1/(\Omega\Delta{T})\gg 1$.  The same factor for the spin frequency
metric components is simply equal to the ratio of the short and long
observation lengths $\Delta{T}_{\text{s}}/\Delta{T}_{\text{l}}$ and
for templates in the orbital angular frequency dimension the same
ratio is given by
$\Omega\tau(\Delta{T}_{\text{s}}/\Delta{T}_{\text{l}}$.  This final
result gives the false impression that by using short coherent
segments one is able to better determine the orbital angular frequency
of a source than by using long segments.  We stress here that template
placement is related to, but not equivalent to, parameter estimation.
Template placement using the metric defined on fractional \ac{SNR}
($\propto$ the log-likelihood) loss determines how finely a parameter
space must be sampled in order to avoid missing a potential signal.
Parameter estimation makes use of the likelihood itself and it's
accuracy improves with \ac{SNR}.

%
We mention at this point that the reader may question our choice of
phase model parameterization.  It is true that parameterizations like
$x=a\cos(\Omega t_{\mathrm{asc}}),y=a\sin(\Omega t_{\mathrm{asc}})$
would potentially simplify the metric calculations and that a rescaling
of the frequency parameter could do the same.  However, since a
general ``fix-all'' re-parameterization eludes the author, it was
decided that the physical parameters were justifiable choices.  The
fact that the metric elements are functions of the parameters
themselves is problematic from an implementation sense since this will
potentially make template placement more complicated.  Conversely, the
parameter choice leads to a clearer understanding of how the
semi-coherent metric scales with the unambiguous physical parameters
of the system.

%
In conclusion, in deriving the semi-coherent metric for known binary
systems we have provided a necessary tool for the continued
sensitivity improvement of searches for continuously emitting sources
of \ac{EM} and \ac{GW} emission.  With focus on \ac{GW} searches the
advanced detector~\cite{2010CQGra..27h4006H} era is fast approaching and \acp{LMXB} are an ever
more realistic detection target.  The optimality of search sensitivity
at fixed computational cost has been approximately achieved by the
Einstein@Home \ac{GW} searches~\cite{2009PhRvD..80d2003A} for unknown
isolated \acp{NS}.  The achievement has arisen through the application
of semi-coherent search strategies described
in~\cite{2009PhRvL.103r1102P} and are similar to those outlined in
this work.  It is our hope that search strategies based on the
semi-coherent methods described here will be employed in searches for
the prime \ac{GW} \ac{LMXB} targets such as Sco X-1 and Cyg X-2.  

%
\begin{acknowledgments}
  We would like to thank Karl Wette, Evan Goetz, Holger Pletsch, Reinhard Prix,
  and the LIGO Scientific Collaboration continuous waves working group
  for many useful discussions.
\end{acknowledgments}
%
\appendix
\section{Re-parameterized phase coordinates\label{app:ucoords}}
In Section~\ref{sec:shortcoherent} there is reference to a
re-parameterization of the phase approximation given by
Eq.~\ref{eq:loweccentricityphasemodel} into the $\bm{u}$ coordinates
used in Eq.~\ref{eq:uphase}.  The $\bm{u}^{(m)}$ coordinates for
the $m$'th segment are mapped onto the physical parameters
$\bm{\theta}$ via   
\begin{widetext}
\begin{eqnarray}
  \Phi^{(m)}_{0} &=& \frac{\Phi_0}{2\pi} + \nu (t +
  t^{(n)}_{\mathrm{mid}} - t_{\mathrm{ref}})\nonumber \\
  &&- a \nu \left(\frac{\eta}{2} (3 + \cos[2 (t - t_{\mathrm{asc}} + t^{(n)}_{\mathrm{mid}}) \Omega]) - 
    \sin[(t - t_{\mathrm{asc}} + t^{(n)}_{\mathrm{mid}}) \Omega] - \frac{\kappa}{2} \sin[2 (t - t_{\mathrm{asc}} + t^{(n)}_{\mathrm{mid}}) \Omega]\right), \\
   u^{(m)}_{1} &=& \nu + 
 a \nu \Omega(\cos[(t - t_{\mathrm{asc}} + t^{(n)}_{\mathrm{mid}}) \Omega] + 
    \kappa \cos[2 (t - t_{\mathrm{asc}} + t^{(n)}_{\mathrm{mid}}) \Omega] + 
    \eta \sin[2 (t - t_{\mathrm{asc}} + t^{(n)}_{\mathrm{mid}}) \Omega]), \\
  u^{(m)}_{2} &=& a \nu  \Omega^2\left(2 \eta \cos[2 (t - t_{\mathrm{asc}} + t^{(n)}_{\mathrm{mid}}) \Omega] - 
   \sin[(t - t_{\mathrm{asc}} + t^{(n)}_{\mathrm{mid}}) \Omega] - 
   2 \kappa \sin[2 (t - t_{\mathrm{asc}} + t^{(n)}_{\mathrm{mid}}) \Omega]\right), \\
  u^{(m)}_{3} &=& -a \nu \Omega^3 \left(\cos[(t - t_{\mathrm{asc}} + t^{(n)}_{\mathrm{mid}}) \Omega] +
   4 \kappa \cos[2 (t - t_{\mathrm{asc}} + t^{(n)}_{\mathrm{mid}}) \Omega] + 
   4 \eta \sin[2 (t - t_{\mathrm{asc}} +
     t^{(n)}_{\mathrm{mid}}) \Omega]\right),\\
   u^{(m)}_{4} &=&-a \nu \Omega^4 \left(8 \eta \cos[2 (t - t_{\mathrm{asc}} + t^{(n)}_{\mathrm{mid}}) \Omega] - 
   \sin[(t - t_{\mathrm{asc}} + t^{(n)}_{\mathrm{mid}}) \Omega] - 8 \kappa \sin[2 (t - t_{\mathrm{asc}} + t^{(n)}_{\mathrm{mid}})
     \Omega]\right).
\end{eqnarray}
\end{widetext}
Note that we give expressions only up to $4^{\mathrm{th}}$ order in
time.  If higher orders are required then it is likely that the
coherent observation time spans a large fraction (or more) of a single
orbital period.  In this regime using the Taylor-expanded phase may
not be a sensible choice and it may be more practical to use the fully
coherent metric on the physical parameters.  
%
\section{Coherent metric elements for long ($\Delta T \gg P$) segment lengths\label{app:coherentmetricelements}}

To leading order in $\kappa$ and $\eta$ (where appropriate) the elements of the coherent metric on the
physical parameters space
$\bm{\theta}=\left\{\nu,a,t_{\mathrm{asc}},\Omega,\kappa,\eta\right\}$
in the limit $\Delta T\gg P$ and
$\Omega\Delta{T}\gg 1$ are
\begin{widetext}
\begin{eqnarray}
  g_{\nu\nu} &=& \frac{\pi^{2}\Delta{T}^{2}}{3} \\
  g_{aa} &=& 2\pi ^2\nu^{2} \left(1+\frac{e^{2}}{4}\right)\\
  g_{t_{\mathrm{asc}}t_{\mathrm{asc}}} &=& 2 \pi^{2}\nu^{2}a^2\Omega^{2} \left(1+e^{2}\right) \\
  g_{\Omega\Omega} &=&
  \frac{1}{6}\pi^{2}\nu^{2}a^{2}\Delta{T}^{2}\left(1+e^{2}\right)\\
  g_{\kappa\kappa} &=& \frac{1}{2} a^2 \pi ^2 \nu ^2\\
  g_{\eta\eta} &=& \frac{1}{2} a^2 \pi ^2 \nu ^2\\
  g_{\nu{a}} = g_{a\nu} &=& 
  2\pi^{2}\nu\Bigg\{{a}\left(1+\frac{e^{2}}{4}\right)
  - \frac{2}{\Omega}\cos\left(\frac{\Delta{T} \Omega
    }{2}\right) \cos\left(\Omega
    t_{\mathrm{asc}}\right)+\frac{1}{\Omega}\cos\left(\Omega\Delta
      {T} \right) \Big[\eta  \sin\left(2 \Omega
      t_{\mathrm{asc}}\right)- \kappa  \cos\left(2 \Omega
      t_{\mathrm{asc}}\right)\Big]\Bigg\} \\
    g_{\nu{t_{\mathrm{asc}}}} = g_{{t_{\mathrm{asc}}}\nu} &=& 4 a \pi ^2 \nu  \left\{\cos\left(\frac{\Delta{T} \Omega }{2}\right) \sin\left(\Omega  t_{\text{asc}}\right)+\frac{1}{2}\cos\left(\Delta{T} \Omega \right) \Big[\eta  \cos\left[2 \Omega  t_{\text{asc}}\right]+\kappa  \sin\left(2 \Omega  t_{\text{asc}}\right)\Big]\right\}\\
  g_{\nu\Omega} = g_{\Omega\nu} &=& \frac{2\pi ^2\nu{a}\Delta{T}}{\Omega}
    \left\{\cos\left(\Omega  t_{\mathrm{asc}}\right)
    \sin\left(\frac{\Delta{T} \Omega }{2}\right)
    + \frac{1}{2}
    \sin\left(\Omega\Delta {T}\right)\Big[\kappa\cos\left(2\Omega{t}_{\mathrm{asc}}\right)-\eta\sin\left(2\Omega{t}_{\mathrm{asc}}\right)\Big]\right\}\\
  g_{\nu\kappa} = g_{\kappa\nu} &=& -\frac{\pi ^2 \nu{a}}{\Omega}
    \cos\left(\Omega\Delta {T} \right) \cos\left(2 \Omega
      t_{\mathrm{asc}}\right)+\frac{1}{2}\pi^{2}\nu {a}^2
  \kappa \\
  g_{\nu\eta} = g_{\eta\nu} &=& \frac{\pi ^2 \nu{a}}{\Omega}
    \cos\left(\Omega\Delta {T} \right) \sin\left(2 \Omega
      t_{\mathrm{asc}}\right)+\frac{1}{2} \pi^{2}\nu{a}^2 \eta
  \\
 g_{a{t_{\mathrm{asc}}}} = g_{t_{\mathrm{asc}}a} &=& 0 \\ 
  g_{a\Omega} = g_{\Omega{a}} &=& -\frac{\pi ^2 \nu ^2{a}}{\Omega}\Bigg\{
    \cos(\Omega\Delta{T} ) \cos\left(2 \Omega
      t_{\mathrm{asc}}\right)+ \kappa\cos\left(\frac{\Omega\Delta {T}
      }{2}\right) \Bigg[\cos\left(\Omega
        t_{\mathrm{asc}}\right)+\cos\left(3 \Omega
        t_{\mathrm{asc}}\right)\Big[1-2 \cos\left(\Omega\Delta {T}
      \right)\Big] \Bigg]\nonumber \\
  &&+ 2\eta\cos\left(\frac{\Omega\Delta {T}
        }{2}\right) \sin\left(\Omega  t_{\mathrm{asc}}\right) \Bigg[\cos\left(\Omega\Delta {T}
      \right)\Big[1+2\cos\left(2\Omega
          t_{\mathrm{asc}}\right)\Big]-\cos\left(2 \Omega
        t_{\mathrm{asc}}\right)-1\Bigg]
    \Bigg\} \\
  g_{a\kappa} = g_{\kappa{a}} &=& \frac{1}{2} \pi ^2\nu^{2}{a} \kappa\\
  g_{a\eta} = g_{\eta{a}} &=& \frac{1}{2}\pi ^2\nu^{2}{a} \eta\\
  g_{t_{\mathrm{asc}}\Omega} = g_{{\Omega}t_{\mathrm{asc}}} &=& \pi^{2}\nu^{2}a^2\Bigg\{
  \cos\left(\Omega\Delta {T}\right)
    \sin\left(2 \Omega  t_{\mathrm{asc}}\right)+2 \Omega
    t_{\mathrm{asc}} \nonumber\\
  &&+\frac{4\eta}{3} \cos\left(\frac{\Omega\Delta
        {T}}{2}\right) \Big[3 \cos\left(\Omega
      t_{\mathrm{asc}}\right)+\cos\left(3 \Omega  t_{\mathrm{asc}}\right) \left(2 \cos\left(\Omega\Delta {T}
    \right)-1\right) \Big] 
  \nonumber \\
  &&+\frac{4\kappa}{3} \cos\left(\frac{\Omega\Delta
        {T}}{2}\right) \Big[3 \sin\left(\Omega
      t_{\mathrm{asc}}\right)+\sin\left(3 \Omega  t_{\mathrm{asc}}\right) \left(2 \cos\left(\Omega\Delta {T}
    \right)-1\right) \Big] \Bigg\} \\
  g_{t_{\mathrm{asc}}\kappa} = g_{{\kappa}t_{\mathrm{asc}}} &=& -\pi ^2\nu^{2}a^{2}\Omega \eta \\
 g_{t_{\mathrm{asc}}\eta} = g_{{\eta}t_{\mathrm{asc}}} &=& \pi ^2\nu ^2a^{2} \Omega\kappa\\
  g_{\Omega\kappa} = g_{\kappa\Omega} &=& -\frac{\pi^{2}\nu^{2}a^2}{\Omega}\Bigg\{
    \frac{1}{3}\cos\left(\frac{\Omega\Delta {T}}{2}\right) \Big[3
      \cos\left(\Omega  t_{\mathrm{asc}}\right)+\cos\left(3\Omega{t}_{\mathrm{asc}}\right)\left(2
      \cos\left(\Omega\Delta {T}\right)-1\right) \Big] \nonumber\\
  &&+ \frac{\eta}{4}\Big[4\Omega{t}_{\mathrm{asc}}-\cos\left(2
    \Omega\Delta {T}\right) \sin\left(4\Omega{t}_{\mathrm{asc}}\right)\Big]
  +\frac{\kappa}{4}\cos\left(2 \Omega\Delta {T}\right) \cos\left(4\Omega{t}_{\mathrm{asc}}\right)
    \Bigg\}\\
  g_{\Omega\eta} = g_{\eta\Omega} &=& \frac{\pi^{2}\nu^{2}a^2}{\Omega}\Bigg\{
    \frac{1}{3}\cos\left(\frac{\Omega\Delta {T}}{2}\right) \Big[3
      \sin\left(\Omega  t_{\mathrm{asc}}\right)+\sin\left(3\Omega{t}_{\mathrm{asc}}\right)\left(2
      \cos\left(\Omega\Delta {T}\right)-1\right) \Big] \nonumber\\
  &&+ \frac{\kappa}{4}\Big[4\Omega{t}_{\mathrm{asc}}+\cos\left(2
    \Omega\Delta {T}\right) \sin\left(4\Omega{t}_{\mathrm{asc}}\right)\Big]
  +\frac{\eta}{4}\cos\left(2 \Omega\Delta {T}\right) \cos\left(4\Omega{t}_{\mathrm{asc}}\right)
    \Bigg\}\\
  g_{\kappa\eta} = g_{\eta\kappa} &=& 0
\end{eqnarray}
\end{widetext}
%

\section{Semi-coherent metric elements\label{app:metricelements}}

Here we give the expressions for the metric elements in the eccentric
orbit case for a phase model defined by Eq.~\ref{eq:loweccentricityphasemodel}
on the set of parameters $\bm{\theta}=\left\{\nu,a,t_{\mathrm{asc}},\Omega,\kappa,\eta\right\}$.
The following elements are valid in the limits of
$e\ll 1$ and for $\tau\gg \Delta{T}$ and $\tau\gg P$,
\begin{widetext}
\begin{eqnarray}
  G_{\nu\nu} &=& \frac{\pi^2\Delta{T}^{2}}{3} - \frac{\pi^2a^2}{12\Omega^{2}\Delta{T}^{2}} \left(48+3 e^{2}-24 \Delta{T}^2 \Omega^2-6e^{2}\Delta{T}^2 \Omega^2-48 \cos(\Delta{T} \Omega)-3e^{2} \cos(2 \Delta{T} \Omega)\right)\\
 G_{aa} &=& 2\pi^{2}\nu^{2}\left\{1+\frac{e^{2}}{4}-\frac{2}{\Omega^{2}\Delta{T}^{2}}\left[1-\cos(\Omega
    \Delta{T})+\frac{e^{2}}{16}-\frac{e^{2}}{16}\cos(2\Omega \Delta{T})\right]\right\}
  \\
  G_{t_{\mathrm{asc}}t_{\mathrm{asc}}} &=& 2\pi^{2}\nu^{2}a^{2}\Omega^{2}\left\{1+e^{2}-\frac{2}{\Omega^{2}\Delta{T}^{2}}\left[1-\cos(\Omega
    \Delta{T})+\frac{e^{2}}{4}-\frac{e^{2}}{4}\cos(2\Omega \Delta{T})\right]\right\} \\
  G_{\Omega\Omega} &=& \frac{1}{6}a^{2}\pi^{2}\nu^{2}\Delta
  \Delta{T}^{2}\left\{1+e^{2}-\frac{1}{2\Omega^{2}\Delta{T}^{2}}\left[4-4\cos(\Omega
    \Delta{T})+e^{2}-e^{2}\cos(2\Omega \Delta{T})\right]\right\} \\
  G_{\kappa\kappa} &=& 2a^2 \nu^2 \pi^2 \left[1 -
  \frac{\left(1-\cos\left(2\Omega
    \Delta{T}\right)\right)}{8\Omega^{2}\Delta{T}^{2}}\right] \\
  G_{\eta\eta} &=& 2a^2 \nu^2 \pi^2 \left[1 -
  \frac{\left(1-\cos\left(2\Omega
    \Delta{T}\right)\right)}{8\Omega^{2}\Delta{T}^{2}}\right] \\
  G_{\nu{a}} &=& G_{a\nu} = 2\pi^{2}\nu\left\{1+\frac{e^{2}}{4}-\frac{2}{\Omega^{2}\Delta{T}^{2}}\left[1-\cos(\Omega
    \Delta{T})+\frac{e^{2}}{16}-\frac{e^{2}}{16}\cos(2\Omega \Delta{T})\right]\right\}\\
  G_{\nu{t_{\mathrm{asc}}}} &=& G_{{t_{\mathrm{asc}}}\nu} = 0\\
  G_{\nu\Omega} &=& G_{\Omega\nu} = \frac{4\pi^{2}\nu{a}^{2}}{\Omega^{2}\Delta{T}^{2}}\left\{1-\cos\left(\Omega{\Delta{T}}\right)+\frac{e^{2}}{16}-\frac{e^{2}}{16}\cos\left(2\Omega{\Delta{T}}\right)-\frac{e^{2}\Omega{\Delta{T}}}{16}\sin\left(2\Omega{\Delta{T}}\right)\right\}\\
  G_{\nu\kappa} &=& G_{\kappa\nu} = \frac{1}{2}\pi^{2}\nu a^{2}\kappa\left\{1 - \frac{1}{2\Omega^{2}\Delta{T}^{2}}\left[1-\cos\left(\Omega{\Delta{T}}\right)\right]\right\}\\
  G_{\nu\eta} &=& G_{\eta\nu} = \frac{1}{2}\pi^{2}\nu a^{2}\eta\left\{1 -
  \frac{1}{2\Omega^{2}\Delta{T}^{2}}\left[1-\cos\left(\Omega{\Delta{T}}\right)\right]\right\}\\
  G_{a{t_{\mathrm{asc}}}} &=& G_{t_{\mathrm{asc}}a} = 0 \\
  G_{a\Omega} &=& G_{\Omega{a}} = \frac{4\pi^{2}a\nu^{2}}{\Omega^{2}\Delta{T}^{2}}\left\{1-\cos\left(\Omega{\Delta{T}}\right)-\frac{\Omega{\Delta{T}}}{2}\sin\left(\Omega{\Delta{T}}\right)+\frac{e^{2}}{16}+\frac{e^{2}}{16}\cos\left(2\Omega{\Delta{T}}\right)-\frac{e^{2}\Omega{\Delta{T}}}{16}\sin\left(2\Omega{\Delta{T}}\right)\right\}\\
  G_{a\kappa} &=& G_{\kappa{a}} = \frac{1}{2}\pi^{2}\nu^{2}a\kappa\left\{1-\frac{1}{2\Omega^{2}\Delta{T}^{2}}\left[1-\cos\left(2\Omega{\Delta{T}}\right)\right]\right\}\\
  G_{a\eta} &=& G_{\eta{a}} =
  \frac{1}{2}\pi^{2}\nu^{2}a\eta\left\{1-\frac{1}{2\Omega^{2}\Delta{T}^{2}}\left[1-\cos\left(2\Omega{\Delta{T}}\right)\right]\right\}\\
  G_{t_{\mathrm{asc}}\Omega} &=& G_{{\Omega}t_{\mathrm{asc}}} = 2a^{2}\pi^{2}\nu^{2}t_{\mathrm{asc}}\left\{1+e^{2}-\frac{2}{\Omega^{2}\Delta{T}^{2}}\left[1+\frac{e^{2}}{4}-\cos\left(\Omega{\Delta{T}}\right)-\frac{e^{2}}{4}\cos\left(2\Omega{\Delta{T}}\right)\right]\right\}\\
  G_{t_{\mathrm{asc}}\kappa} &=& G_{{\kappa}t_{\mathrm{asc}}} = -\pi^{2}\nu^{2}a^{2}\eta\left\{1-\frac{1}{2\Omega^{2}\Delta{T}^{2}}\left[1-\cos\left(2\Omega{\Delta{T}}\right)\right]\right\}\\
  G_{t_{\mathrm{asc}}\eta} &=& G_{{\eta}t_{\mathrm{asc}}} = \pi^{2}\nu^{2}a^{2}\kappa\left\{1-\frac{1}{2\Omega^{2}\Delta{T}^{2}}\left[1-\cos\left(2\Omega{\Delta{T}}\right)\right]\right\}\\
  G_{\Omega\kappa} &=& G_{\kappa\Omega} = -\frac{a^2 \nu^2 \pi ^2}{4\Omega^{3}\Delta{T}^{2}} \left(-\kappa-2 \eta t_{\mathrm{asc}} \Omega+4 \eta \Delta{T}^2 t_{\mathrm{asc}} \Omega^3+(\kappa+2 \eta t_{\mathrm{asc}} \Omega) \cos(2 \Delta{T} \Omega)+\kappa \Delta{T} \Omega \sin(2 \Delta{T} \Omega)\right)\\
  G_{\Omega\eta} &=& G_{\eta\Omega} = \frac{a^2 \nu^2 \pi ^2}{4\Omega^{3}\Delta{T}^{2}} \left(\eta-2 \kappa t_{\mathrm{asc}} \Omega+4 \kappa \Delta{T}^2 t_{\mathrm{asc}} \Omega^3-(\eta-2 \kappa t_{\mathrm{asc}} \Omega) \cos(2 \Delta{T} \Omega)-\eta \Delta{T} \Omega \sin(2 \Delta{T} \Omega)\right)\\
  G_{\kappa\eta} &=& G_{\eta\kappa} = 0.
\end{eqnarray}
The following metric elements are the Taylor-expanded versions, to
leading order in $\Omega \Delta{T}$ and in $e$, of those given above
\begin{eqnarray}
  G^{\mathrm{T}}_{\nu\nu} &=& \frac{\pi^2\Delta{T}^{2}}{3}\\
  G^{\mathrm{T}}_{aa} &=&  \frac{1}{6}\pi^{2}\nu^{2}\Omega^{2}\Delta{T}^{2} \\
  G^{\mathrm{T}}_{t_{\mathrm{asc}}t_{\mathrm{asc}}} &=& \frac{1}{6}\pi^{2}\nu^{2}a^{2}\Omega^{2}\Delta{T}^{2} \\
  G^{\mathrm{T}}_{\Omega\Omega} &=&
  \frac{1}{72}\pi^{2}\nu^{2}a^{2}\Omega^{2}\Delta{T}^{2}\tau^{2} \\
  G^{\mathrm{T}}_{\kappa\kappa} &=& \frac{1}{6}a^2 \nu^2 \pi^2 \Omega^{2}\Delta{T}^{2}\\
  G^{\mathrm{T}}_{\eta\eta} &=& \frac{1}{6}a^2 \nu^2 \pi^2 \Omega^{2}\Delta{T}^{2}\\
  G^{\mathrm{T}}_{\nu{a}} = G^{\mathrm{T}}_{a\nu} &=& \frac{1}{6}\pi^{2}a\nu\Omega^{2}\Delta{T}^{2}\\
  G^{\mathrm{T}}_{\nu{t_{\mathrm{asc}}}} =
  G^{\mathrm{T}}_{t_{\mathrm{asc}}\nu} &=& 0\\
  G^{\mathrm{T}}_{\nu\Omega} = G^{\mathrm{T}}_{\Omega\nu} &=& \frac{1}{6}\pi^{2}a^{2}\nu\Omega{\Delta{T}}^{2}\\
  G^{\mathrm{T}}_{\nu\kappa} = G^{\mathrm{T}}_{\kappa\nu} &=& \frac{1}{6}\pi^{2}a^{2}\nu\kappa\Omega^{2}{\Delta{T}}^{2}\\
  G^{\mathrm{T}}_{\nu\eta} = G^{\mathrm{T}}_{\eta\nu} &=&
  \frac{1}{6}\pi^{2}a^{2}\nu\eta\Omega^{2}{\Delta{T}}^{2}\\
  G^{\mathrm{T}}_{a{t_{\mathrm{asc}}}} =
  G^{\mathrm{T}}_{t_{\mathrm{asc}}a} &=& 0\\
  G^{\mathrm{T}}_{a\Omega} = G^{\mathrm{T}}_{\Omega{a}} &=& \frac{1}{6}\pi^{2}\nu^{2}a\Omega{\Delta{T}}^{2}\\
  G^{\mathrm{T}}_{a\kappa} = G^{\mathrm{T}}_{\kappa{a}} &=& \frac{1}{6}\pi^{2}\nu^{2}a\kappa\Omega^{2}\Delta{T}^{2}\\
  G^{\mathrm{T}}_{a\eta} = G^{\mathrm{T}}_{\eta{a}} &=&
  \frac{1}{6}\pi^{2}\nu^{2}a\eta\Omega^{2}\Delta{T}^{2}\\
  G^{\mathrm{T}}_{t_{\mathrm{asc}}\Omega} =
  G^{\mathrm{T}}_{\Omega{t_{\mathrm{asc}}}} &=& \frac{1}{6}\pi^{2}\nu^{2}a^{2}t_{\mathrm{asc}}\Omega^{3}\Delta{T}^{2}\\
  G^{\mathrm{T}}_{t_{\mathrm{asc}}\kappa} =
  G^{\mathrm{T}}_{\kappa{t_{\mathrm{asc}}}} &=& -\frac{1}{3}\pi^{2}\nu^{2}a^{2}\eta\Omega^{3}\Delta{T}^{2}\\
  G^{\mathrm{T}}_{t_{\mathrm{asc}}\eta} =
  G^{\mathrm{T}}_{\eta{t_{\mathrm{asc}}}} &=& \frac{1}{3}\pi^{2}\nu^{2}a^{2}\kappa\Omega^{3}\Delta{T}^{2}\\
  G^{\mathrm{T}}_{\Omega\kappa} = G^{\mathrm{T}}_{\kappa\Omega} &=&
  \frac{1}{6}\pi^{2}\nu^{2}a^{2}\Delta{T}^{2}\left(\kappa\Omega-2\eta t_{\mathrm{asc}}\Omega^{2}\right)\\
  G^{\mathrm{T}}_{\Omega\eta} = G^{\mathrm{T}}_{\eta\Omega} &=&
  \frac{1}{6} \left(a^2 \eta \nu^2 \pi^2 \Omega+2 a^2 \kappa \nu^2
  \pi^2 t_{\mathrm{asc}}  \Omega^2\right) \Delta{T}^2\\
  G^{\mathrm{T}}_{\kappa\eta} = G^{\mathrm{T}}_{\eta\kappa} &=& 0. 
\end{eqnarray}
\end{widetext}
The metric given in Eq.~\ref{eq:semi-coherentmetric} is a further
approximation in which the off-diagonal terms have been set to zero.
This approximation is valid due to the fact that these terms cause only minor mismatch ellipse
rotations in the physical parameter space.  It follows that the
diagonal elements themselves are a very good approximation to the
eigenvalues of the metric and the eigenvectors are very well aligned
with the physical dimensions of the parameter space. 

\bibliography{masterbib}

\begin{thebibliography}{35}
\expandafter\ifx\csname natexlab\endcsname\relax\def\natexlab#1{#1}\fi
\expandafter\ifx\csname bibnamefont\endcsname\relax
  \def\bibnamefont#1{#1}\fi
\expandafter\ifx\csname bibfnamefont\endcsname\relax
  \def\bibfnamefont#1{#1}\fi
\expandafter\ifx\csname citenamefont\endcsname\relax
  \def\citenamefont#1{#1}\fi
\expandafter\ifx\csname url\endcsname\relax
  \def\url#1{\texttt{#1}}\fi
\expandafter\ifx\csname urlprefix\endcsname\relax\def\urlprefix{URL }\fi
\providecommand{\bibinfo}[2]{#2}
\providecommand{\eprint}[2][]{\url{#2}}

\bibitem[{\citenamefont{{Anderson} et~al.}(1990)\citenamefont{{Anderson},
  {Gorham}, {Kulkarni}, {Prince}, and {Wolszczan}}}]{1990Natur.346...42A}
\bibinfo{author}{\bibfnamefont{S.~B.} \bibnamefont{{Anderson}}},
  \bibinfo{author}{\bibfnamefont{P.~W.} \bibnamefont{{Gorham}}},
  \bibinfo{author}{\bibfnamefont{S.~R.} \bibnamefont{{Kulkarni}}},
  \bibinfo{author}{\bibfnamefont{T.~A.} \bibnamefont{{Prince}}},
  \bibnamefont{and}
  \bibinfo{author}{\bibfnamefont{A.}~\bibnamefont{{Wolszczan}}},
  \bibinfo{journal}{nat} \textbf{\bibinfo{volume}{346}}, \bibinfo{pages}{42}
  (\bibinfo{year}{1990}).

\bibitem[{\citenamefont{{Anderson}}(1992)}]{1992PhDT........20A}
\bibinfo{author}{\bibfnamefont{S.~R.} \bibnamefont{{Anderson}}}, Ph.D. thesis,
  \bibinfo{school}{Illinois Inst.~of Tech., Chicago.} (\bibinfo{year}{1992}).

\bibitem[{\citenamefont{{Faulkner} et~al.}(2004)\citenamefont{{Faulkner},
  {Stairs}, {Kramer}, {Lyne}, {Hobbs}, {Possenti}, {Lorimer}, {Manchester},
  {McLaughlin}, {D'Amico} et~al.}}]{2004MNRAS.355..147F}
\bibinfo{author}{\bibfnamefont{A.~J.} \bibnamefont{{Faulkner}}},
  \bibinfo{author}{\bibfnamefont{I.~H.} \bibnamefont{{Stairs}}},
  \bibinfo{author}{\bibfnamefont{M.}~\bibnamefont{{Kramer}}},
  \bibinfo{author}{\bibfnamefont{A.~G.} \bibnamefont{{Lyne}}},
  \bibinfo{author}{\bibfnamefont{G.}~\bibnamefont{{Hobbs}}},
  \bibinfo{author}{\bibfnamefont{A.}~\bibnamefont{{Possenti}}},
  \bibinfo{author}{\bibfnamefont{D.~R.} \bibnamefont{{Lorimer}}},
  \bibinfo{author}{\bibfnamefont{R.~N.} \bibnamefont{{Manchester}}},
  \bibinfo{author}{\bibfnamefont{M.~A.} \bibnamefont{{McLaughlin}}},
  \bibinfo{author}{\bibfnamefont{N.}~\bibnamefont{{D'Amico}}},
  \bibnamefont{et~al.}, \bibinfo{journal}{mnras}
  \textbf{\bibinfo{volume}{355}}, \bibinfo{pages}{147} (\bibinfo{year}{2004}),
  \eprint{arXiv:astro-ph/0408228}.

\bibitem[{\citenamefont{{Ransom} et~al.}(2003)\citenamefont{{Ransom}, {Cordes},
  and {Eikenberry}}}]{2003ApJ...589..911R}
\bibinfo{author}{\bibfnamefont{S.~M.} \bibnamefont{{Ransom}}},
  \bibinfo{author}{\bibfnamefont{J.~M.} \bibnamefont{{Cordes}}},
  \bibnamefont{and} \bibinfo{author}{\bibfnamefont{S.~S.}
  \bibnamefont{{Eikenberry}}}, \bibinfo{journal}{apj}
  \textbf{\bibinfo{volume}{589}}, \bibinfo{pages}{911} (\bibinfo{year}{2003}),
  \eprint{arXiv:astro-ph/0210010}.

\bibitem[{\citenamefont{{Hessels} et~al.}(2007)\citenamefont{{Hessels},
  {Ransom}, {Stairs}, {Kaspi}, and {Freire}}}]{2007ApJ...670..363H}
\bibinfo{author}{\bibfnamefont{J.~W.~T.} \bibnamefont{{Hessels}}},
  \bibinfo{author}{\bibfnamefont{S.~M.} \bibnamefont{{Ransom}}},
  \bibinfo{author}{\bibfnamefont{I.~H.} \bibnamefont{{Stairs}}},
  \bibinfo{author}{\bibfnamefont{V.~M.} \bibnamefont{{Kaspi}}},
  \bibnamefont{and} \bibinfo{author}{\bibfnamefont{P.~C.~C.}
  \bibnamefont{{Freire}}}, \bibinfo{journal}{apj}
  \textbf{\bibinfo{volume}{670}}, \bibinfo{pages}{363} (\bibinfo{year}{2007}),
  \eprint{0707.1602}.

\bibitem[{\citenamefont{{Ransom} et~al.}(2005)\citenamefont{{Ransom},
  {Hessels}, {Stairs}, {Kaspi}, {Freire}, and {Backer}}}]{2005ASPC..328..199R}
\bibinfo{author}{\bibfnamefont{S.}~\bibnamefont{{Ransom}}},
  \bibinfo{author}{\bibfnamefont{J.}~\bibnamefont{{Hessels}}},
  \bibinfo{author}{\bibfnamefont{I.}~\bibnamefont{{Stairs}}},
  \bibinfo{author}{\bibfnamefont{V.}~\bibnamefont{{Kaspi}}},
  \bibinfo{author}{\bibfnamefont{P.}~\bibnamefont{{Freire}}}, \bibnamefont{and}
  \bibinfo{author}{\bibfnamefont{D.}~\bibnamefont{{Backer}}}, in
  \emph{\bibinfo{booktitle}{Binary Radio Pulsars}}, edited by
  \bibinfo{editor}{\bibnamefont{{F.~A.~Rasio \& I.~H.~Stairs}}}
  (\bibinfo{year}{2005}), vol. \bibinfo{volume}{328} of
  \emph{\bibinfo{series}{Astronomical Society of the Pacific Conference
  Series}}, pp. \bibinfo{pages}{199--+}.

\bibitem[{\citenamefont{{Messenger} and {Woan}}(2007)}]{2007CQGra..24..469M}
\bibinfo{author}{\bibfnamefont{C.}~\bibnamefont{{Messenger}}} \bibnamefont{and}
  \bibinfo{author}{\bibfnamefont{G.}~\bibnamefont{{Woan}}},
  \bibinfo{journal}{Classical and Quantum Gravity}
  \textbf{\bibinfo{volume}{24}}, \bibinfo{pages}{469} (\bibinfo{year}{2007}),
  \eprint{arXiv:gr-qc/0703155}.

\bibitem[{\citenamefont{{Wood} et~al.}(1991)\citenamefont{{Wood}, {Norris},
  {Hertz}, {Vaughan}, {Michelson}, {Mitsuda}, {Lewin}, {van Paradijs},
  {Penninx}, and {van der Klis}}}]{1991ApJ...379..295W}
\bibinfo{author}{\bibfnamefont{K.~S.} \bibnamefont{{Wood}}},
  \bibinfo{author}{\bibfnamefont{J.~P.} \bibnamefont{{Norris}}},
  \bibinfo{author}{\bibfnamefont{P.}~\bibnamefont{{Hertz}}},
  \bibinfo{author}{\bibfnamefont{B.~A.} \bibnamefont{{Vaughan}}},
  \bibinfo{author}{\bibfnamefont{P.~F.} \bibnamefont{{Michelson}}},
  \bibinfo{author}{\bibfnamefont{K.}~\bibnamefont{{Mitsuda}}},
  \bibinfo{author}{\bibfnamefont{W.~H.~G.} \bibnamefont{{Lewin}}},
  \bibinfo{author}{\bibfnamefont{J.}~\bibnamefont{{van Paradijs}}},
  \bibinfo{author}{\bibfnamefont{W.}~\bibnamefont{{Penninx}}},
  \bibnamefont{and} \bibinfo{author}{\bibfnamefont{M.}~\bibnamefont{{van der
  Klis}}}, \bibinfo{journal}{apj} \textbf{\bibinfo{volume}{379}},
  \bibinfo{pages}{295} (\bibinfo{year}{1991}).

\bibitem[{\citenamefont{{Vaughan} et~al.}(1994)\citenamefont{{Vaughan}, {van
  der Klis}, {Wood}, {Norris}, {Hertz}, {Michelson}, {van Paradijs}, {Lewin},
  {Mitsuda}, and {Penninx}}}]{1994ApJ...435..362V}
\bibinfo{author}{\bibfnamefont{B.~A.} \bibnamefont{{Vaughan}}},
  \bibinfo{author}{\bibfnamefont{M.}~\bibnamefont{{van der Klis}}},
  \bibinfo{author}{\bibfnamefont{K.~S.} \bibnamefont{{Wood}}},
  \bibinfo{author}{\bibfnamefont{J.~P.} \bibnamefont{{Norris}}},
  \bibinfo{author}{\bibfnamefont{P.}~\bibnamefont{{Hertz}}},
  \bibinfo{author}{\bibfnamefont{P.~F.} \bibnamefont{{Michelson}}},
  \bibinfo{author}{\bibfnamefont{J.}~\bibnamefont{{van Paradijs}}},
  \bibinfo{author}{\bibfnamefont{W.~H.~G.} \bibnamefont{{Lewin}}},
  \bibinfo{author}{\bibfnamefont{K.}~\bibnamefont{{Mitsuda}}},
  \bibnamefont{and}
  \bibinfo{author}{\bibfnamefont{W.}~\bibnamefont{{Penninx}}},
  \bibinfo{journal}{ApJ} \textbf{\bibinfo{volume}{435}}, \bibinfo{pages}{362}
  (\bibinfo{year}{1994}).

\bibitem[{\citenamefont{et~al.}(2007{\natexlab{a}})}]{2007PhRvD..76h2001A}
\bibinfo{author}{\bibfnamefont{L.~S. C. B.~A.} \bibnamefont{et~al.}},
  \bibinfo{journal}{prd} \textbf{\bibinfo{volume}{76}}, \bibinfo{pages}{082001}
  (\bibinfo{year}{2007}{\natexlab{a}}), \eprint{arXiv:gr-qc/0605028}.

\bibitem[{\citenamefont{et~al.}(2007{\natexlab{b}})}]{2007PhRvD..76h2003A}
\bibinfo{author}{\bibfnamefont{L.~S. C. B.~A.} \bibnamefont{et~al.}},
  \bibinfo{journal}{prd} \textbf{\bibinfo{volume}{76}}, \bibinfo{pages}{082003}
  (\bibinfo{year}{2007}{\natexlab{b}}), \eprint{arXiv:astro-ph/0703234}.

\bibitem[{\citenamefont{{Dhurandhar} and
  {Vecchio}}(2001)}]{2001PhRvD..63l2001D}
\bibinfo{author}{\bibfnamefont{S.~V.} \bibnamefont{{Dhurandhar}}}
  \bibnamefont{and}
  \bibinfo{author}{\bibfnamefont{A.}~\bibnamefont{{Vecchio}}},
  \bibinfo{journal}{prd} \textbf{\bibinfo{volume}{63}}, \bibinfo{pages}{122001}
  (\bibinfo{year}{2001}), \eprint{arXiv:gr-qc/0011085}.

\bibitem[{\citenamefont{{Watts} et~al.}(2008)\citenamefont{{Watts}, {Krishnan},
  {Bildsten}, and {Schutz}}}]{2008MNRAS.389..839W}
\bibinfo{author}{\bibfnamefont{A.~L.} \bibnamefont{{Watts}}},
  \bibinfo{author}{\bibfnamefont{B.}~\bibnamefont{{Krishnan}}},
  \bibinfo{author}{\bibfnamefont{L.}~\bibnamefont{{Bildsten}}},
  \bibnamefont{and} \bibinfo{author}{\bibfnamefont{B.~F.}
  \bibnamefont{{Schutz}}}, \bibinfo{journal}{mnras}
  \textbf{\bibinfo{volume}{389}}, \bibinfo{pages}{839} (\bibinfo{year}{2008}),
  \eprint{0803.4097}.

\bibitem[{\citenamefont{Balasubramanian
  et~al.}(1996)\citenamefont{Balasubramanian, Sathyaprakash, and
  Dhurandhar}}]{bala96:_gravit_binaries_metric}
\bibinfo{author}{\bibfnamefont{R.}~\bibnamefont{Balasubramanian}},
  \bibinfo{author}{\bibfnamefont{B.~S.} \bibnamefont{Sathyaprakash}},
  \bibnamefont{and} \bibinfo{author}{\bibfnamefont{S.~V.}
  \bibnamefont{Dhurandhar}}, \bibinfo{journal}{Phys. Rev. D}
  \textbf{\bibinfo{volume}{53}}, \bibinfo{pages}{3033} (\bibinfo{year}{1996}).

\bibitem[{\citenamefont{Owen}(1996)}]{owen96:_search_templates}
\bibinfo{author}{\bibfnamefont{B.~J.} \bibnamefont{Owen}},
  \bibinfo{journal}{Phys. Rev. D} \textbf{\bibinfo{volume}{53}},
  \bibinfo{pages}{6749} (\bibinfo{year}{1996}).

\bibitem[{\citenamefont{{Owen} and
  {Sathyaprakash}}(1999)}]{1999PhRvD..60b2002O}
\bibinfo{author}{\bibfnamefont{B.~J.} \bibnamefont{{Owen}}} \bibnamefont{and}
  \bibinfo{author}{\bibfnamefont{B.~S.} \bibnamefont{{Sathyaprakash}}},
  \bibinfo{journal}{Phys. Rev. D} \textbf{\bibinfo{volume}{60}},
  \bibinfo{pages}{022002} (\bibinfo{year}{1999}).

\bibitem[{\citenamefont{{Brady} and {Creighton}}(2000)}]{2000PhRvD..61h2001B}
\bibinfo{author}{\bibfnamefont{P.~R.} \bibnamefont{{Brady}}} \bibnamefont{and}
  \bibinfo{author}{\bibfnamefont{T.}~\bibnamefont{{Creighton}}},
  \bibinfo{journal}{\prd} \textbf{\bibinfo{volume}{61}},
  \bibinfo{pages}{082001} (\bibinfo{year}{2000}), \eprint{arXiv:gr-qc/9812014}.

\bibitem[{\citenamefont{{Pletsch}}(2008)}]{2008PhRvD..78j2005P}
\bibinfo{author}{\bibfnamefont{H.~J.} \bibnamefont{{Pletsch}}},
  \bibinfo{journal}{\prd} \textbf{\bibinfo{volume}{78}},
  \bibinfo{pages}{102005} (\bibinfo{year}{2008}), \eprint{0807.1324}.

\bibitem[{\citenamefont{{Pletsch} and {Allen}}(2009)}]{2009PhRvL.103r1102P}
\bibinfo{author}{\bibfnamefont{H.~J.} \bibnamefont{{Pletsch}}}
  \bibnamefont{and} \bibinfo{author}{\bibfnamefont{B.}~\bibnamefont{{Allen}}},
  \bibinfo{journal}{Physical Review Letters} \textbf{\bibinfo{volume}{103}},
  \bibinfo{pages}{181102} (\bibinfo{year}{2009}), \eprint{0906.0023}.

\bibitem[{\citenamefont{{Blandford} and
  {Teukolsky}}(1976)}]{1976ApJ...205..580B}
\bibinfo{author}{\bibfnamefont{R.}~\bibnamefont{{Blandford}}} \bibnamefont{and}
  \bibinfo{author}{\bibfnamefont{S.~A.} \bibnamefont{{Teukolsky}}},
  \bibinfo{journal}{\apj} \textbf{\bibinfo{volume}{205}}, \bibinfo{pages}{580}
  (\bibinfo{year}{1976}).

\bibitem[{\citenamefont{{Roy}}(1988)}]{1988ormo.book.....R}
\bibinfo{author}{\bibfnamefont{A.~E.} \bibnamefont{{Roy}}},
  \emph{\bibinfo{title}{{Orbital motion}}} (\bibinfo{publisher}{Institute of
  Physics, 2005}, \bibinfo{year}{1988}).

\bibitem[{\citenamefont{{Lange} et~al.}(2001)\citenamefont{{Lange}, {Camilo},
  {Wex}, {Kramer}, {Backer}, {Lyne}, and {Doroshenko}}}]{2001MNRAS.326..274L}
\bibinfo{author}{\bibfnamefont{C.}~\bibnamefont{{Lange}}},
  \bibinfo{author}{\bibfnamefont{F.}~\bibnamefont{{Camilo}}},
  \bibinfo{author}{\bibfnamefont{N.}~\bibnamefont{{Wex}}},
  \bibinfo{author}{\bibfnamefont{M.}~\bibnamefont{{Kramer}}},
  \bibinfo{author}{\bibfnamefont{D.~C.} \bibnamefont{{Backer}}},
  \bibinfo{author}{\bibfnamefont{A.~G.} \bibnamefont{{Lyne}}},
  \bibnamefont{and}
  \bibinfo{author}{\bibfnamefont{O.}~\bibnamefont{{Doroshenko}}},
  \bibinfo{journal}{mnras} \textbf{\bibinfo{volume}{326}}, \bibinfo{pages}{274}
  (\bibinfo{year}{2001}), \eprint{arXiv:astro-ph/0102309}.

\bibitem[{\citenamefont{{Neyman} and {Pearson}}(1933)}]{NeymanPearson1933}
\bibinfo{author}{\bibfnamefont{J.}~\bibnamefont{{Neyman}}} \bibnamefont{and}
  \bibinfo{author}{\bibfnamefont{E.}~\bibnamefont{{Pearson}}},
  \bibinfo{journal}{Philosophical Transactions of the Royal Society of London}
  \textbf{\bibinfo{volume}{231}}, \bibinfo{pages}{289} (\bibinfo{year}{1933}).

\bibitem[{\citenamefont{{Searle}}(2008)}]{2008arXiv0804.1161S}
\bibinfo{author}{\bibfnamefont{A.~C.} \bibnamefont{{Searle}}},
  \bibinfo{journal}{ArXiv e-prints}  (\bibinfo{year}{2008}),
  \eprint{0804.1161}.

\bibitem[{\citenamefont{{Jaranowski} et~al.}(1998)\citenamefont{{Jaranowski},
  {Kr{\'o}lak}, and {Schutz}}}]{1998PhRvD..58f3001J}
\bibinfo{author}{\bibfnamefont{P.}~\bibnamefont{{Jaranowski}}},
  \bibinfo{author}{\bibfnamefont{A.}~\bibnamefont{{Kr{\'o}lak}}},
  \bibnamefont{and} \bibinfo{author}{\bibfnamefont{B.~F.}
  \bibnamefont{{Schutz}}}, \bibinfo{journal}{\prd}
  \textbf{\bibinfo{volume}{58}}, \bibinfo{pages}{063001}
  (\bibinfo{year}{1998}), \eprint{arXiv:gr-qc/9804014}.

\bibitem[{\citenamefont{{Wijnands} et~al.}(2008)\citenamefont{{Wijnands},
  {Altamirano}, {Soleri}, {Degenaar}, {Rea}, {Casella}, {Patruno}, and
  {Linares}}}]{2008AIPC.1068.....W}
\bibinfo{editor}{\bibfnamefont{R.}~\bibnamefont{{Wijnands}}},
  \bibinfo{editor}{\bibfnamefont{D.}~\bibnamefont{{Altamirano}}},
  \bibinfo{editor}{\bibfnamefont{P.}~\bibnamefont{{Soleri}}},
  \bibinfo{editor}{\bibfnamefont{N.}~\bibnamefont{{Degenaar}}},
  \bibinfo{editor}{\bibfnamefont{N.}~\bibnamefont{{Rea}}},
  \bibinfo{editor}{\bibfnamefont{P.}~\bibnamefont{{Casella}}},
  \bibinfo{editor}{\bibfnamefont{A.}~\bibnamefont{{Patruno}}},
  \bibnamefont{and}
  \bibinfo{editor}{\bibfnamefont{M.}~\bibnamefont{{Linares}}}, eds.,
  \emph{\bibinfo{title}{{A Decade of Accreting Millisecond X-Ray Pulsars}}},
  vol. \bibinfo{volume}{1068} of \emph{\bibinfo{series}{American Institute of
  Physics Conference Series}} (\bibinfo{year}{2008}).

\bibitem[{\citenamefont{{Prix}}(2007{\natexlab{a}})}]{2007PhRvD..75b3004P}
\bibinfo{author}{\bibfnamefont{R.}~\bibnamefont{{Prix}}},
  \bibinfo{journal}{\prd} \textbf{\bibinfo{volume}{75}},
  \bibinfo{pages}{023004} (\bibinfo{year}{2007}{\natexlab{a}}),
  \eprint{arXiv:gr-qc/0606088}.

\bibitem[{\citenamefont{{Conway} and {Sloane}}(1993)}]{CONWAYSLOANE}
\bibinfo{author}{\bibfnamefont{J.~H.} \bibnamefont{{Conway}}} \bibnamefont{and}
  \bibinfo{author}{\bibfnamefont{N.~J.~A.} \bibnamefont{{Sloane}}},
  \emph{\bibinfo{title}{{Sphere Packings, Lattices, and Groups}}}
  (\bibinfo{publisher}{{New York: Springer-Verlag}}, \bibinfo{year}{1993}),
  \bibinfo{edition}{{second}} ed.

\bibitem[{\citenamefont{{Prix}}(2007{\natexlab{b}})}]{2007CQGra..24..481P}
\bibinfo{author}{\bibfnamefont{R.}~\bibnamefont{{Prix}}},
  \bibinfo{journal}{Class. Quant. Grav.} \textbf{\bibinfo{volume}{24}},
  \bibinfo{pages}{S481} (\bibinfo{year}{2007}{\natexlab{b}}).

\bibitem[{\citenamefont{{Messenger} et~al.}(2009)\citenamefont{{Messenger},
  {Prix}, and {Papa}}}]{2009PhRvD..79j4017M}
\bibinfo{author}{\bibfnamefont{C.}~\bibnamefont{{Messenger}}},
  \bibinfo{author}{\bibfnamefont{R.}~\bibnamefont{{Prix}}}, \bibnamefont{and}
  \bibinfo{author}{\bibfnamefont{M.~A.} \bibnamefont{{Papa}}},
  \bibinfo{journal}{\prd} \textbf{\bibinfo{volume}{79}},
  \bibinfo{pages}{104017} (\bibinfo{year}{2009}), \eprint{0809.5223}.

\bibitem[{\citenamefont{Harry et~al.}(2008)\citenamefont{Harry, Fairhurst, and
  Sathyaprakash}}]{harry-2008}
\bibinfo{author}{\bibfnamefont{I.~W.} \bibnamefont{Harry}},
  \bibinfo{author}{\bibfnamefont{S.}~\bibnamefont{Fairhurst}},
  \bibnamefont{and} \bibinfo{author}{\bibfnamefont{B.~S.}
  \bibnamefont{Sathyaprakash}}, \bibinfo{journal}{Class. Quant. Grav.}
  \textbf{\bibinfo{volume}{25}}, \bibinfo{pages}{184027}
  (\bibinfo{year}{2008}).

\bibitem[{\citenamefont{Babak}(2008)}]{babak-2008}
\bibinfo{author}{\bibfnamefont{S.}~\bibnamefont{Babak}},
  \bibinfo{journal}{Class. Quant. Grav.} \textbf{\bibinfo{volume}{25}},
  \bibinfo{pages}{195011} (\bibinfo{year}{2008}).

\bibitem[{\citenamefont{{Manca} and {Vallisneri}}(2010)}]{2010PhRvD..81b4004M}
\bibinfo{author}{\bibfnamefont{G.~M.} \bibnamefont{{Manca}}} \bibnamefont{and}
  \bibinfo{author}{\bibfnamefont{M.}~\bibnamefont{{Vallisneri}}},
  \bibinfo{journal}{\prd} \textbf{\bibinfo{volume}{81}},
  \bibinfo{pages}{024004} (\bibinfo{year}{2010}), \eprint{0909.0563}.

\bibitem[{\citenamefont{{Harry} and {the LIGO Scientific
  Collaboration}}(2010)}]{2010CQGra..27h4006H}
\bibinfo{author}{\bibfnamefont{G.~M.} \bibnamefont{{Harry}}} \bibnamefont{and}
  \bibinfo{author}{\bibnamefont{{the LIGO Scientific Collaboration}}},
  \bibinfo{journal}{Classical and Quantum Gravity}
  \textbf{\bibinfo{volume}{27}}, \bibinfo{pages}{084006}
  (\bibinfo{year}{2010}).

\bibitem[{\citenamefont{{Abbott} et~al.}(2009)\citenamefont{{Abbott}, {Abbott},
  {Adhikari}, {Ajith}, {Allen}, {Allen}, {Amin}, {Anderson}, {Anderson},
  {Arain} et~al.}}]{2009PhRvD..80d2003A}
\bibinfo{author}{\bibfnamefont{B.~P.} \bibnamefont{{Abbott}}},
  \bibinfo{author}{\bibfnamefont{R.}~\bibnamefont{{Abbott}}},
  \bibinfo{author}{\bibfnamefont{R.}~\bibnamefont{{Adhikari}}},
  \bibinfo{author}{\bibfnamefont{P.}~\bibnamefont{{Ajith}}},
  \bibinfo{author}{\bibfnamefont{B.}~\bibnamefont{{Allen}}},
  \bibinfo{author}{\bibfnamefont{G.}~\bibnamefont{{Allen}}},
  \bibinfo{author}{\bibfnamefont{R.~S.} \bibnamefont{{Amin}}},
  \bibinfo{author}{\bibfnamefont{S.~B.} \bibnamefont{{Anderson}}},
  \bibinfo{author}{\bibfnamefont{W.~G.} \bibnamefont{{Anderson}}},
  \bibinfo{author}{\bibfnamefont{M.~A.} \bibnamefont{{Arain}}},
  \bibnamefont{et~al.}, \bibinfo{journal}{\prd} \textbf{\bibinfo{volume}{80}},
  \bibinfo{pages}{042003} (\bibinfo{year}{2009}), \eprint{0905.1705}.

\end{thebibliography}

\end{document}